%% file: notes.tex
\documentclass[a4paper,11pt,english]{article}
\usepackage[utf8]{inputenc}
\usepackage[T1]{fontenc}
\usepackage{lmodern, amsmath, adjustbox, a4wide, multirow, xcolor, graphicx, xfrac, slashed, cancel,mathtools}
\usepackage[numbers, sort&compress, elide]{natbib}
\usepackage{jheppub}

\makeatletter\g@addto@macro\bfseries{\boldmath}\makeatother

\def\figureautorefname~#1\null{fig.\,#1\null}

\newcommand{\subappref}[1]{\hyperref[#1]{appendix~\ref{#1}}}

\def\equationautorefname~#1\null{eq.\,(#1)\null}

\usepackage{tikz}
\usetikzlibrary{positioning}
\newcommand{\minitab}[2][c]{\begin{tabular}{@{}#1@{}}#2\end{tabular}}

\newcommand{\myline}{\\[2mm]\hline\noalign{\vskip2mm}}

\newcommand*{\bs}[1]{\boldsymbol{#1}}
\newcommand*{\anb}[1]{\langle#1\rangle}
\newcommand*{\asb}[1]{\langle#1]}
\newcommand*{\sab}[1]{[#1\rangle}
\newcommand*{\sqb}[1]{[#1]}
\newcommand*{\Anb}[1]{\anb{\bs{#1}}}
\newcommand*{\Asb}[1]{\asb{\bs{#1}}}
\newcommand*{\Sab}[1]{\sab{\bs{#1}}}
\newcommand*{\Sqb}[1]{\sqb{\bs{#1}}}
\newcommand*{\tdp}[2]{\tilde{s}_{#1#2}}
\newcommand{\amp}[4]{($#1${}$#2${}$#3${}$#4$)}
\newcommand*{\psic}{\psi^c}
\DeclareMathOperator{\Tr}{Tr}
\DeclareMathOperator{\ceil}{ceil}

\allowdisplaybreaks[4]
\newcommand{\beqa}{\begin{eqnarray}}
\newcommand{\eeqa}{\end{eqnarray}}
\newcommand{\beq}{\begin{equation}}
\newcommand{\eeq}{\end{equation}}

\author[a]{Gauthier Durieux,}
\author[a,b,c]{Teppei Kitahara,}
\author[d]{Camila S.\ Machado,}
\author[a]{Yael Shadmi,}
\author[a]{and Yaniv Weiss}

\affiliation[a]{Physics Department, Technion---Israel Institute of Technology,\\Technion city, Haifa 3200003, Israel}
\affiliation[b]{Institute for Advanced Research, Nagoya University,\\Furo-cho Chikusa-ku, Nagoya 464-8601, Japan}
\affiliation[c]{Kobayashi-Maskawa Institute for the Origin of Particles and the Universe, Nagoya University, Furo-cho Chikusa-ku, Nagoya 464-8602, Japan}
\affiliation[d]{PRISMA$^+$  Cluster of Excellence \& Mainz Institute for Theoretical Physics, Johannes Gutenberg-Universit\"at Mainz, 55099 Mainz, Germany}

\title{Constructing massive on-shell contact terms}

\abstract{%
The purely on-shell approach to effective field theories requires the construction of independent contact terms.
Employing the little-group-covariant massive-spinor formalism, we present the first systematic derivation of independent four-point contact terms involving massive scalars, spin-$1/2$ fermions, and vectors.
Independent three-point amplitudes are also listed for massive particles up to spin-$3$.
We make extensive use of the simple relations between massless and massive amplitudes in this formalism.
Our general results are specialized to the (broken-phase) particle content of the electroweak sector of the standard model.
The (anti)symmetrization among identical particles is then accounted for.
This work opens the way for the on-shell computation of massive four-point amplitudes.
}

\preprint{MITP/20-046}

\begin{document}
\sloppy 

\makeatletter\renewcommand{\@fpheader}{\ }\makeatother
\maketitle

\section{Introduction}

On-shell methods find a natural application in the framework of effective field theories (EFTs), which provide a model-independent description of particle interactions, positing just the low-energy particle content and symmetries.
In the context of the standard-model effective field theory, they have for instance been used to elucidate the patterns observed in the renormalization of dimension-six operators~\cite{Cheung:2015aba} and in their interference with the standard model~\cite{Azatov:2016sqh}, which are somewhat obscured in the Lagrangian approach, and to derive anomalous dimensions beyond leading order in the loop and EFT expansions~\cite{Bern:2019wie, Bern:2020ikv, Jiang:2020mhe, EliasMiro:2020tdv}.
Many of these applications only involve massless amplitudes, for which on-shell techniques are especially powerful.
EFTs describing massive particles are of particular importance however.
At energies probed experimentally, the masses of standard-model particles are often relevant and the interactions of the heaviest ones may hold the key to a better understanding of electroweak symmetry breaking.
New particles postulated beyond the standard model are moreover typically massive.
Whether these masses are sub-eV or above TeV, their effects are often important.
The little-group-covariant massive-spinor formalism of ref.~\cite{Arkani-Hamed:2017jhn} (see also ref.~\cite{Conde:2016vxs}) facilitates the study of these theories from an on-shell perspective.

Beyond the applications described above, the on-shell program allows for an alternative description of general EFTs, substituting the classification and enumeration of independent operators by that of independent on-shell amplitudes~\cite{Henning:2015daa, Henning:2017fpj, Shadmi:2018xan, Henning:2019enq, Henning:2019mcv, Henning:2017fpj, Aoude:2019tzn, Durieux:2019eor, Durieux:2019siw, Ma:2019gtx, Falkowski:2019zdo}.
A general $n$-point tree-level amplitude can in principle be sequentially derived from three-point amplitudes, as well as four- and higher-point contact terms.
Both renormalizable and non-renomalizable three-point amplitudes must first be identified.
These can be \emph{glued} together to form the factorizable parts of tree-level four-point amplitudes whose analytic structure is characterized by single poles due to internal propagators~\cite{Arkani-Hamed:2017jhn, Bern:1994zx}.
For recursion relations involving massive particles, see e.g.\ refs.~\cite{Cohen:2010mi,Franken:2019wqr, Falkowski:2020aso}.
The remainder of the four-point amplitude consists of pole-free non-factorizable contact terms which map to higher-dimensional operators in a Lagrangian formulation.
Their classification therefore replaces that of EFT operators contributing to the four-point amplitude of interest.
Repeating this process to form higher-point amplitudes and isolating their local contact terms, one can characterize general EFTs and construct their amplitudes up to a given dimension.
For massive amplitudes, a general procedure for the construction of contact terms is lacking.
In this paper, we take a first step in this direction, deriving bases for four-point non-factorizable contact terms in the massive-spinor formalism of ref.~\cite{Arkani-Hamed:2017jhn}.

We cover general three-point amplitudes for spins up to $3$, and four-point amplitudes involving scalars, spin-$1/2$ fermions, and vectors.
Our results extend and complete several previous analyses.
Four-point contact terms featuring a massive scalar or vector plus gluons were classified up to dimension-13 in ref.~\cite{Shadmi:2018xan}.
The massive three-point amplitudes generated by bosonic dimension-six standard-model operators were presented in ref.~\cite{Aoude:2019tzn}.
Several four-point contact terms required to restore unitarity were also identified in ref.~\cite{Aoude:2019tzn} (see also ref.~\cite{Bachu:2019ehv}).
Renormalizable~\cite{Christensen:2018zcq} and non-renormalizable~\cite{Durieux:2019eor} contributions to general three-point amplitudes featuring particle of the standard-model electroweak sector were classified.
Reference~\cite{Durieux:2019eor} also derived the general four-point fermion-fermion-vector-scalar contact terms.
Massless contact terms were classified in ref.~\cite{Durieux:2019siw} for spins $0,1/2,1,2$ and up to dimension-eight.
An alternative classification of independent massless amplitudes in terms of momentum twistors was presented in ref.~\cite{Falkowski:2019zdo}.

We will discuss three types of bases:
a. the \emph{spinor-structure} bases comprising the basic independent building blocks which span generic amplitudes.
These replace the standard independent Lorentz invariants constructed out of external polarizations;
b. the analogous \emph{stripped-contact-term} (SCT) bases, comprising the basic building blocks required for spanning the pole-free non-factorizable amplitudes;
and c. the \emph{contact-term} bases which span the full non-factorizable amplitudes.
The latter are relevant for EFT classification and on-shell computations.
Throughout, we exploit the massless limits of massive spinor structures.
In the formalism of ref.~\cite{Arkani-Hamed:2017jhn}, the spinors associated with massive particles are  bolded, to imply symmetrization over little-group indices, while those of massless particles remain unbolded.
The massless limit can then be obtained, in most cases, by unbolding the massive structures~\cite{Arkani-Hamed:2017jhn}.
In fact, the reverse process, which we refer to as \emph{bolding}, yields an elegant derivation of spinor-structure bases.
These are obtained by starting with a judicious choice of independent massless amplitudes and covariantizing them with respect to the massive little group.

We apply our results to derive the general four-point amplitudes of the standard-model electroweak sector, for one generation, without imposing SU(2)$_L\times$U(1)$_Y$.
We list all contributions of dimension $\leq6$, as well as several other contributions of interest, which only arise at higher dimensions. 
The symmetrization over identical electroweak bosons greatly constrains the minimal dimension at which certain structures can first appear.
Since the standard-model gauge symmetry is not imposed, the contact-term bases we derive can be used to parametrize general EFT extensions of the standard model, including both strongly-coupled theories, with the EFT scale near the electroweak scale, and weakly coupled extensions, featuring hierarchically-separated scales, with the Higgs VEV $v\ll\Lambda$. 
The results furthermore automatically include the full $v/\Lambda$ expansion, thus encoding the geometric interpretation of generic standard-model EFTs in the Lagrangian formulation~\cite{Alonso:2015fsp,Alonso:2016oah,Helset:2020yio,Cohen:2020xca}.

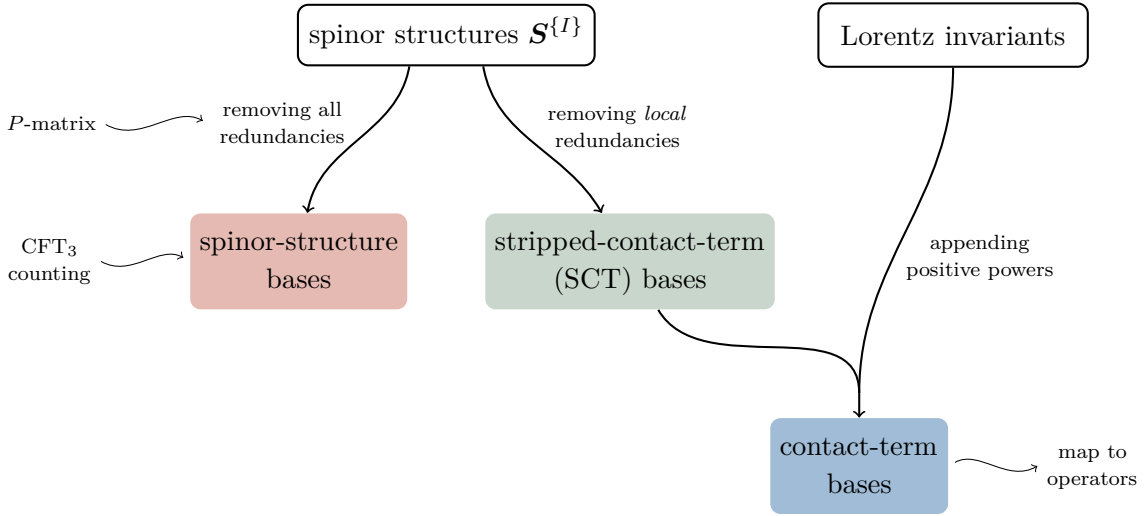
\begin{figure}[t]
\centering
\definecolor{mred}{rgb}{0.896,  0.728,  0.7}
\definecolor{mblue}{rgb}{0.628,  0.736,  0.84}
\definecolor{mgreen}{rgb}{0.788,  0.84 ,  0.8}
\adjustbox{max width=\textwidth}{\begin{tikzpicture}
\node (ss) [rounded corners, thick, inner sep=1.5mm, draw] {\minitab{spinor structures $\bs{S}^{\{I\}}$}};
\node (li) [right=3cm of ss, thick, rounded corners, inner sep=3mm, draw] {Lorentz invariants};
\node (ss_b)  [below left =of ss, yshift=-1cm, xshift= 25mm, rounded corners, fill=mred, fill opacity=1, text opacity=1] {\minitab{spinor-structure\\bases}};
\node (sct_b) [below right=of ss, yshift=-1cm, xshift=-25mm, rounded corners, fill=mgreen,   fill opacity=1, text opacity=1] {\minitab{stripped-contact-term\\ (SCT) bases}};
\draw [->, thick] (ss.south)++(-5mm,0) to[out=-100, in=80] node (redun) [above left =-2mm, pos=.6, font=\scriptsize]{\minitab{removing all\\ redundancies}} (ss_b);
\draw [->, thick] (ss.south)++(5mm,0) to[out=-80, in=120] node          [above right=-2mm, pos=.6, font=\scriptsize]{\minitab{removing \emph{local}\\redundancies}} (sct_b);
\node (w) [below right=of sct_b]{};
\draw [thick] (sct_b) to[out=-60, in=90] (w.center);
\draw [thick] (li) to[out=-90, in=90] node[below right=-2mm, font=\scriptsize]{\minitab{appending\\ positive powers}} (w.center);
\node (ct_b) [below=2mm of w, rounded corners, fill=mblue, fill opacity=1, text opacity=1]{\minitab{contact-term\\bases}};
\draw [->, thick] (w.center) to[out=-90, in=90]  (ct_b);
\draw [->, shorten <=1mm] (ct_b.east)  to[out=30,in=-150] ++( 1.2cm,0) node[right, font=\scriptsize]{\minitab{map to\\ operators}};
\draw [<-, shorten <=1mm] (ss_b.west)  to[out=150,in=-30] ++(-1.2cm,0) node[left,  font=\scriptsize]{\minitab{CFT$_3$\\counting}};
\draw [<-, shorten <=1mm] (redun.west) to[out=150,in=-30] ++(-1.4cm,  0) node[left,  font=\scriptsize]{\minitab{$P$-matrix}};
\end{tikzpicture}}
\caption{%
A schematic description of our analysis.
A set of spinor structures can be reduced to a \emph{spinor-structure} basis, whose dimension is determined by counting 3d conformal field theory (CFT$_3$) correlators.
The $P$-matrix of inner-products is useful in this process, and all redundancies are removed.  
Alternatively, the set of spinor structures can be reduced to a \emph{stripped-contact-term} (SCT) basis, by eliminating \emph{local} linear combinations, namely those with prefactors containing only positive powers of the Lorentz invariants and masses.
This is appropriate for the construction of manifestly-local contact terms.
Finally, the \emph{contact-term} basis, which corresponds to an operator basis, is obtained by appending positive powers of the independent Lorentz invariants to the elements of the SCT basis, and dropping newly redundant combinations.
}
\label{fig:map}
\end{figure}

The basics of our approach are outlined in \autoref{sec:basics}.
We then turn to a classification of the independent spinor structures contributing to three- and four-point amplitudes, without imposing any symmetry requirements apart from Lorentz, and treating all particles as distinguishable.
\hyperref[sec:3pt]{Section~\ref{sec:3pt}} addresses three-point amplitudes.
In \autoref{sec:4pt}, we study four-point amplitudes, starting with a quick review of the massless case, and proceeding to amplitudes involving massive scalars, spin-$1/2$ fermions and vectors.
These general results are then specialized to the particle content of the electroweak sector of the standard model in \autoref{sec:ew_application}.
At this stage, the (anti)symmetrization required by spin statistics is in particular accounted for, while colour and flavour structures are left implicit.
A roadmap of the different basis constructions and their inter-relations is sketched in \autoref{fig:map}.

\section{Basics}
\label{sec:basics}
We use the little-group-covariant massive-spinor formalism~\cite{Arkani-Hamed:2017jhn}.
A massive amplitude carries $2s_i$ symmetrized little-group indices for each external particle of spin $s_i$.
This symmetrization is implicit in the boldface notation of ref.~\cite{Arkani-Hamed:2017jhn}.
We use $i,j,\ldots$ to label external particles, and $I,J,\ldots=1,2$, to denote massive little-group indices.
Apart from \autoref{sec:ew_application}, we treat external particles as distinguishable.

The non-factorizable part of the amplitude can be written as a sum,
\beq\label{eq:basis}
\sum_n \bs{S}_n^{\{I\}}\, L_n[\tdp{i}{j},\epsilon(p_i,p_j,p_k,p_l)]
\eeq
where $\{I\}$ collectively denotes the little-group indices of all external particles, while $L_n$ is a polynomial of the Lorentz invariants $\tdp{i}{j}=2p_i\cdot p_j$ and, for five or more external legs, of the antisymmetric contractions of four momenta $\epsilon(p_i,p_j,p_k,p_l) \equiv \epsilon_{\mu\nu\rho\sigma}p_i^\mu p_j^\nu p_k^\rho p_l^\sigma$.
Each spinor structure $\bs{S}_n^{\{I\}}$ is the minimal product of angle and square spinor brackets, possibly containing momentum insertions, obtained by \emph{stripping} the contact term off all Lorentz invariants.
For clarity, we always consider manifestly local contact terms which involve no negative power of Lorentz invariants (or spinor products, in the massless case).
Our main objective is the construction of contact-term bases, spanning \autoref{eq:basis}.
As an intermediate step, we identify the sets of independent spinor structures $\{\bs{S}^{\{I\}}\}$ from which contact-term bases can be generated by multiplying each element $\bs{S}^{\{I\}}$ by Lorentz invariants.
We refer to these sets as bases of \emph{stripped contact terms} (SCTs).
For the purpose of constructing an SCT basis, a spinor structure is redundant if it can be expressed as a linear combination of the other structures, with prefactors involving only \emph{non-negative} powers of Lorentz invariants 
and masses (in the massive case).
Locality can therefore be kept manifest at all stages.
From an EFT point of view, forbidding negative powers of dimensionful quantities also avoids removing lower-dimensional operators in favour of higher-dimensional combinations.
Inverse powers of the masses are only required in the $\bs i \rangle [ \bs i /m_i$ combinations appearing in the polarization vectors of particles of spin-$1$ and higher.

Another basis of interest is the \emph{spinor-structure basis}, which can be used to span a generic amplitude.
Here a different notion of redundancy is appropriate, with no restriction imposed on the prefactors.
In particular, the prefactors  generically take the form of rational functions of the invariants in this case.
Bases of SCTs are therefore typically larger than spinor-structure bases.
Note however that an SCT basis cannot be obtained simply by adding some elements to a spinor structure basis (see discussion in \autoref{sec:4pt}).

As we will see, the construction of the spinor-structure bases can directly rely on the massless limit of spinor structures.
Indeed, the transparent relations between massless and massive amplitudes in the spinor formalism~\cite{Arkani-Hamed:2017jhn} will be heavily exploited throughout.
Massless amplitudes are particularly simple.
For each helicity amplitude, the spinor-structure basis collapses to a single element.
All spinor products of identical little-group weight are equal, up to a rational function of Lorentz invariants.
Using the massless relations between spinor products, including Schouten identities, it is straightforward to identify SCT bases for massless amplitudes~\cite{Durieux:2019siw}.

\section{Three-point amplitudes}
\label{sec:3pt}

Formally written in terms of complex momenta, massless three-point amplitudes are unique.
One spinor structure at most is allowed for each helicity configuration:
\begin{equation}
\mathcal{M}_3(h_1,h_2,h_3)\propto
	\begin{cases}
	\,\sqb{12}^{+h_1+h_2-h_3}\: \sqb{23}^{-h_1+h_2+h_3}\: \sqb{13}^{+h_1-h_2+h_3}\,,
		&\text{for }h_1+h_2+h_3>0
		\,,\\
	\anb{12}^{-h_1-h_2+h_3}\anb{23}^{+h_1-h_2-h_3} \anb{13}^{-h_1+h_2-h_3}\,,
		&\text{for }h_1+h_2+h_3<0
		\,.
	\end{cases}
\end{equation}
The spinor structures trivially map to contact terms as all Mandelstam invariants vanish in massless three-point amplitudes.
Non-analyticity (i.e.\ non locality) is allowed in these unphysical objects, and disappears when they are consistently glued into physical four-point amplitudes.

A formula for the construction of massive three-point amplitudes was provided in ref.~\cite{Arkani-Hamed:2017jhn}.
However, the spinor structures it yields are neither mutually independent, nor of minimal mass dimension.
The example of the three-vector $WWZ$ amplitude detailed in ref.~\cite{Durieux:2019eor} is illustrative in this respect.
As in the massless case, independent spinor structures and contact terms are identical since Mandelstam invariants are functions of the masses.
A counting derives from angular momentum conservation.
The number of independent massive three-point spinor structures corresponds to that of irreducible representations in the addition of the three spins~\cite{Costa:2011mg, Durieux:2019eor}.
For $s_1\le s_2\le s_3$ spins, eq.~(4.42) of ref.~\cite{Costa:2011mg} gives:
\begin{equation}
n^\text{3-pt} = (2s_1+1)(2s_2+1) - p(1+p)
\quad\text{with}\quad p\equiv\max\{0,s_1+s_2-s_3\}
\,.\label{eq:n_3pt}
\end{equation}
The number of parity-even and parity-odd structures are respectively found to be $(n^\text{3-pt}+1)/2$ and $(n^\text{3-pt}-1)/2$.

Let us proceed to the systematic construction of those independent spinor structures.
The resulting amplitudes, for spins up to $3$, are listed in \autoref{tab:3pt}.
Similarly to the massless case, one can start from an ansatz involving spinor bilinears.
The little-group constraints require that each $\bs{i}],\bs{i}\rangle$ spinor appears $2s_i$ times.
No distinction is made between square and angle spinors at this point and we denote them jointly as $\bs{i})\equiv \bs{i}]\text{ or }\bs{i}\rangle$.
Schematically, our starting point is thus 
\begin{equation}
(\bs{12})^{s_1+s_2-s_3} (\bs{23})^{-s_1+s_2+s_3} (\bs{13})^{s_1-s_2+s_3} 
= (\bs{12})^{\sum_i s_i - 2 s_3} (\bs{23})^{\sum_i s_i - 2 s_1} (\bs{13})^{\sum_i s_i - 2 s_2} \,.
\label{eq:ansatz_3pt}
\end{equation}
This expression is only possibly sensible when all powers are positive.
This occurs for $\sum_i s_i -2\max_j(s_j)\ge0$, i.e.\ when the highest spin is smaller than the sum of the others.
In our schematic notation, any positive power can be distributed on either angle or square bracket bilinears: $(\bs{ij})^n$ represents any $[\bs{ij}]^k \anb{\bs{ij}}^{n-k}$ product for $n+1$ values of $k=0,...,n$.
The ansatz above thus contains 
\beq\label{eq:number}
\prod_j (\sum_is_i -2 s_j +1)
\eeq
terms in total.
When the highest spin is equal to the sum of the two others, i.e.\ $\sum_i s_i -2\max_j(s_j)=0$, this number actually matches that of independent massive three-point spinor structures given in \autoref{eq:n_3pt}.
For examples of massive three-point amplitudes already worked out explicitly in the literature (e.g.\ in ref.~\cite{Durieux:2019eor}), one can verify that \autoref{eq:ansatz_3pt} then generates the correct independent spinor structures.

The ansatz of \autoref{eq:ansatz_3pt} notably fails when the highest spin is larger than the sum of the two others: $\sum_i s_i -2\max_j(s_j)<0$.
The largest spin is in this case irrelevant to the counting of independent structures in \autoref{eq:n_3pt};
each irreducible representation in the addition of the two smallest spins contributes to the counting by as many units as its dimension;
the sum of these dimensions is however equal to the product of the dimensions of the two lowest spins.
So $n^\text{3-pt} = \prod_i(2s_i+1) \big/(2\max_j( s_j)+1)$ in this case.
The $\sum_i s_i -2\max_j(s_j)<0$ condition is equivalent to the one introduced in ref.~\cite{Durieux:2019siw} indicating that momentum insertions are required to form massless non-factorizable contact terms.
While no momentum insertion was possible in the massless case, one single structure of the $[\bs j(\bs k-\bs l)\bs j\rangle$ form is available in the massive case.
Its use allows to reduce the number of spinors left to be contracted for $\max_j(s_j)$ by an even number.
The remaining contractions are that of a spinor structure with maximal spin $\max(\tilde{s}_i)$ reduced by any integer.
Reducing it until it equals the sum of the other two spins, i.e.\ $\sum_i s_i -2\max(\tilde{s}_i)=0$, the ansatz of \autoref{eq:ansatz_3pt} is again applicable.

For a somewhat more subtle reason, the ansatz of \autoref{eq:ansatz_3pt} also fails when the highest spin is smaller than the sum of the two others: $\sum_i s_i -2\max_j(s_j)>0$.
This only occurs when $\min(s_i)\ge1$.
For example, in the case of three spin-1 massive vectors, one would expect 8 different possibilities:
\begin{align}
(\Anb{12},\Sqb{12})\otimes(\Anb{23},\Sqb{23})\otimes (\Anb{13},\Sqb{13})\,.
\end{align}
It was however shown in ref.~\cite{Durieux:2019eor} that the
\begin{equation}
\begin{aligned}
  &m_1\Anb{12}\Anb{13}\Sqb{23}
  +m_2\Anb{12}\Sqb{13}\Anb{23}
  +m_3\Sqb{12}\Anb{13}\Anb{23}\\
=\:
  &m_1\Sqb{12}\Sqb{13}\Anb{23}
\:+m_2\Sqb{12}\Anb{13}\Sqb{23}
\:+m_3\Anb{12}\Sqb{13}\Sqb{23}
\end{aligned}
\label{eq:rel_vvv}
\end{equation}
equality then applies.
This means that the three-vector amplitude has $8-1=7$ independent structures.
Note \autoref{eq:rel_vvv} can be rewritten as a parity-odd contraction of polarization tensors and momenta $\epsilon(\bs{\varepsilon_1},
\bs{\varepsilon_2},
\bs{\varepsilon_3},
p_{ 1}+p_{ 2}+p_{ 3})$ which vanishes trivially because of momentum conservation.\footnote{Dividing both sides of \autoref{eq:rel_vvv} by $m_1m_2m_3$ and taking the massless limit, one obtains the so-called Dual Ward Identity, or U(1)-decoupling identity~\cite{Mangano:1990by} for the colour-ordered amplitudes, $\mathcal{M}(1^-_{V},2^-_{V},3^+_{V}) = - \mathcal{M}(2^-_{V},1^-_{V},3^+_{V})$ with $\mathcal{M}(1^-_{V},2^-_{V},3^+_{V})  = \anb{12}^3/\anb{13}\anb{23}$.}
It can be multiplied by additional spinors.
One relation is then induced for each possible contraction of these.
They are given by the ansatz of \autoref{eq:ansatz_3pt} for all spins reduced by one unit.
Using \autoref{eq:number}, one counts $\prod_j (\sum_is_i -2 s_j )$ possible contractions.

Eventually, our constructive procedure yields the number of independent massive three-point spinor structures expected from \autoref{eq:n_3pt} and the number of irreducible representations in the spin sum:
\begin{equation}
n^\text{3-pt} =
\left\{
\begin{aligned}
&\frac{\prod_i (2s_i+1)}{2\max_j(s_j)+1}
	&& \text{if } \sum_i s_i -2\max_j(s_j)\le 0\,,
\\
&\prod_j (\sum_is_i -2 s_j +1) - \prod_j (\sum_is_i -2 s_j)
	&& \text{if } \sum_i s_i -2\max_j(s_j)\ge 0\,.
\end{aligned}
\right.
\end{equation}

The assignment of operator dimensions for massive three-point amplitudes is non-trivial and requires a study of their high-energy limit, as performed in ref.~\cite{Durieux:2019eor} for examples relevant to the electroweak sector of the standard model.
Specific combinations of spinor structures are singled out in this process.

\section{Four-point amplitudes}
\label{sec:4pt}

Before addressing the construction of massive amplitudes, let us first recall some properties of massless contact terms.

\subsection{Massless case}
\label{sec:4pt-massless}

In the massless case, spinor structures factorize into products of angle- and square-brackets, since momentum insertions can be decomposed as $[ijk\rangle=\sqb{ij}\anb{jk}$.
Different spinor structures contributing to the same helicity amplitude are related by rational functions of the invariants.
Thus, the spinor-structure basis for each helicity amplitude is one-dimensional.
The SCT basis from which manifestly local contact term can be built is typically larger.
Thus for example, to span the four-fermion \amp++++ amplitude arising at dimension-six, both $\sqb{13}\sqb{24}$ and  $\sqb{14}\sqb{23}$ are needed, even though these structures are related via,
\begin{equation}
\tdp14\,\sqb{13}\sqb{24} + \tdp13 \sqb{14}\sqb{23} =0 \,.
\label{eq:rel_sij}
\end{equation}
At dimension eight however, the sum of $\tdp14\sqb{13}\sqb{24}$ and $\tdp13\sqb{14}\sqb{23}$ is redundant.
The SCT basis in this case therefore has two elements, which can be chosen as  $\sqb{13}\sqb{24}$ and $\sqb{14}\sqb{23}$.
Since there are two independent Lorentz invariants in four-point amplitudes, the full set of contact terms can be obtained from these, multiplied by powers of, for instance, $\tdp13$ and $\tdp14$, and dropping the combination in \autoref{eq:rel_sij} starting at dimension eight.

A systematic construction of massless $n$-point SCTs generated by operators of dimension eight at most was performed in ref.~\cite{Durieux:2019siw}.
We focus here on massless four-point amplitudes, but include all dimensions.
The Schouten identities,
\begin{equation}
\begin{array}{*{5}{c@{\:}}c}
\sqb{12}\sqb{34}	&-&\sqb{13}\sqb{24}	&+&\sqb{14}\sqb{23} &=0\,,\\
\anb{12}\anb{34}	&-&\anb{13}\anb{24}	&+&\anb{14}\anb{23} &=0\,,
\end{array}
\end{equation}
are used to eliminate occurrences of both $\sqb{14}\sqb{23}$ and $\anb{14}\anb{23}$.
We employ the following procedure to account for momentum conservation.
Products of the type $\sqb{ij}\anb{jk}$ ($i\ne k$) are replaced by $-\sqb{il}\anb{lk}$ with $l\ne i,j,k$, iteratively for $j=1,2,3,4$.
The resulting SCT bases for particles of spin-$2$ at most are displayed in \autoref{tab:4pt-massless-sm}, \ref{tab:4pt-massless-three-half}, \ref{tab:4pt-massless-two}, together with the lowest dimension at which each term is generated.
To our knowledge, this is the first time bases involving spin-$3/2$ particles are presented.
For spins higher than one, we use the convention,
\begin{equation}
\dim\{\text{operator}\} =
	n 
	- \sum_i \max(0,\ceil\{s_i-1\}) 
	+ \dim\{\text{spinors}\}
\,,
\label{eq:op_dim_massless}
\end{equation}
where $n$ is the number of external legs.
The full set of contact terms, including higher-dimensional contributions, are obtained by dressing these spinor structures with positive powers of Lorentz invariants, and using relations like \autoref{eq:rel_sij} to remove redundant terms.

\subsection{Massive case}

We now turn to the main part of the paper, namely the construction of massive contact terms.
We focus primarily on four-point amplitudes involving scalars, spin-$1/2$ fermions and vectors, generically denoted $s$, $f$, $v$, respectively.

\subsubsection{Spinor-structure bases}

We begin with an analysis of spinor-structure bases, and discuss two methods for their construction.
One is purely based on massive spinor structures, and the second relies on their massless limits.
The tools we develop are then applied  to the derivation of SCT bases, which is somewhat more involved.

\paragraph{Dimension}

The dimension of the spinor-structure basis in four- and higher-point amplitudes can be derived by counting conformal correlators in three dimensions~\cite{Schomerus:2016epl, Kravchuk:2016qvl,Henning:2017fpj} and is given by
\begin{equation}
n_\text{s}^\text{$\ge$4-pt} = \prod_i(2s_i+1)\,.
\label{eq:n_4pt}
\end{equation}
It is interesting to note that, in the four-point case, this result matches the counting of independent three-point amplitude gluings (see also sec.~5.5 of ref.~\cite{Arkani-Hamed:2017jhn} for an approach similar in spirit).
To see this, introduce a fictitious mediator of spin $S\ge \max\{s_1+s_2,s_3+s_4\}$ and consider the factorization of the four-point $A_4(s_1,s_2,s_3,s_4)$ amplitude on the two $A_3(s_1,s_2,S)$ and $A_3(s_3,s_4,S)$ three-point ones.
As discussed in the previous section, there are $(2s_1+1)(2s_2+1)$ independent $A_3(s_1,s_2,S)$ amplitudes, and $(2s_3+1)(2s_4+1)$ independent $A_3(s_3,s_4,S)$ ones.
So we recover the stated result for the dimension of the spinor-structure basis.
It would be interesting to examine whether such a procedure can be generalized to higher-point amplitudes.

The dimensions of spinor-structure bases in four-point amplitudes featuring scalars, fermions and vectors are thus:
\begin{equation}
\begin{array}{*{9}{c@{\hspace{1.5em}}}c}
            &ssss	& vsss	& ffss	& vvss	& ffvs	& ffff	& vvvs	& vvff	& vvvv\\
n_\text{s}^\text{4-pt}= &1	    & 3	    & 4	    & 9	    & 12	& 16	& 27	& 36	& 81
\end{array}
\,.
\label{e:countingCFT}
\end{equation}
As it also applies to higher-point amplitudes, the counting above implies that no new spinor structure is required in amplitudes featuring additional scalars.
Thus for example, the $ffss$ and $vsss$ spinor structures also form complete sets for $ffsss$ and $vssss$ amplitudes.

\paragraph{Purely massive construction}

Spinor structures of a definite particle content 
form a vector space on the tensor-product space of their little-group indices.
An inner product can therefore be defined as $(\bs S_1,\bs S_2) \equiv \sum_{\{I\}} {\bs{S}_1^{\{I\}}}^* \bs{S}_2^{\{I\}}$.
This can be viewed as the spin-summed interference between two structures, and is a scalar function of the masses, Mandelstam invariants and, for $n>4$, fully antisymmetric $\epsilon(p_{{i}},p_{{j}},p_{{k}},p_{{l}})$ contractions.
Given a set of spinor structures $\{\bs S_m\}$, we can construct the matrix of inner products, or $P$-matrix~\cite{Bonifacio:2018vzv, Boels:2017gyc, Glover:2003cm}, $P_{mn} = (\bs S_m,\bs S_n)$.
All its elements are independent if $\det{P}\neq0$.
A set of $\prod_i(2s_i+1)$ spinor structures with a non-vanishing $P$-matrix determinant therefore forms a basis.
A spinor-structure basis can then be constructed  by iteratively drawing structures from an initial over-complete set.
Such an initial set can easily be formed by multiplying spinor contractions with zero, one, or two momentum insertions (namely, $\Sqb{ij},\Anb{ij}$, $\Asb{ijk}$, and $\Anb{ijki},\Sqb{ijki}$) while ensuring that $2s_i$ spinors appear for each particle of spin $s_i$.
Note that, in four-point amplitudes, structures like $\Anb{ijkl}$ can be reduced using momentum conservation.

$P$-matrix eigenvectors of vanishing eigenvalues yield relations between spinor structures.
When just one structure is redundant, the relation $S_m=\sum_{j\ne m} x_j S_j$ can be obtained by solving the $\sum_{j\ne m} P_{ij} x_j = P_{im}$ linear system for $x_j$.
This method notably allows for an algorithmic derivation of relations like those of \hyperref[eq:rel_vvv]{eqs.\,(\ref{eq:rel_vvv}),}\,\hyperref[eq:rel_ffff]{(\ref{eq:rel_ffff}),}\,\hyperref[eq:rel_ffvs]{(\ref{eq:rel_ffvs}),} and \hyperref[eq:rel_vsss]{(\ref{eq:rel_vsss})}.
It also applies to higher-spin and higher-point amplitudes.
Large systems, where analytical solutions  becomes intractable, can be studied numerically.
Thus for example, relations between spinor structures can be extracted by sampling over masses and momenta,
to fit a sufficiently general ansatz.

\paragraph{Bolding: a massless-based construction}

Massive spinor structures can be classified according to the helicity configuration of their \emph{unbolded} massless counterparts.\footnote{The massive-spinor formalism in principle allows for arbitrary choices of the spin quantization axes.
To make contact with helicity amplitudes in the massless limit, it is natural to choose this to coincide with the particle momentum.
The (non-vanishing) unbolded helicity amplitude is then the leading one obtained in the massless limit.}
The $vvff$ spinor structure $\Anb{12}\Sqb{21}\Sqb{34}$ for instance unbolds to $\tdp12\sqb{34}$, and can therefore be labelled as \amp{0}{0}{\frac{1}2}{\frac{1}2}, or just \amp00++ when there is no ambiguity about the spin of each particle.
We will refer to these labels as \emph{helicity categories}.

A given massive amplitude has exactly $\prod_i(2s_i+1)$ helicity categories.
Since massless spinor structures of different helicities do not interfere, the $P$-matrix associated to any set becomes block-diagonal in the massless limit, with one block per helicity category.
It is therefore straightforward to construct a spinor-structure basis by selecting one representative spinor structure in each helicity category.
The resulting set contains $\prod_i(2s_i+1)$ elements which are guaranteed to be independent, since their $P$-matrix is diagonal for zero masses.
If none of the initial structure vanishes in the massless limit, all diagonal elements are also non-vanishing.
This method applies to higher spins and higher-point amplitudes as well.

Note that the representative spinor structures in each helicity category can moreover be derived from massless amplitudes.
To do so, one simply \emph{bolds} all spinors.
To form massive vector amplitudes, massless scalar amplitudes featuring momentum insertions are required, with the replacement $p_i\to \bs i\rangle[\bs i$.
In this sense, a massive spinor-structure basis can be obtained simply by bolding massless amplitudes.
What is at work behind this magic is of course the little group.
The massless amplitudes constitute certain components of the massive ones.
The bolding amounts to covariantizing them with respect to the full massive little group.

\begin{table}
\scriptsize
\begin{tabular*}{\textwidth}{@{\extracolsep{\fill}}c@{\;}ccc@{\;}cc@{}c}
\\[-1cm]
spins & $n_\text{SCT}$  & $n_\text{s}$
&{hel.\ cat.}
	& {spinor structures}
		& {$n_\text{perm}$}
			& $\min\{d_\text{op}\}$
\myline
$ssss$  & 1 & 1
&\amp0000
	& constant
		& 1	& 4
\myline
$vsss$  & $4\to 3$  & 3
&\amp0000
	& $\Sab{121},\Sab{131}$
		& 1	& 5
\\
&&&\amp+000
	& $\Sqb{1231} \to \Sqb{1231}-\Anb{1231}$
		& $\slashed{2}\to1$
				& 7
\myline
$ffss$  & 4 & 4
&\amp++00
	& $\Sqb{12}$
		& 2	& 5
\\
&&&\amp+-00
	& $\Sab{132}$
		& 2	& 6
\myline
$vvss$  & $10\to9$  & 9
&\amp0000
	& $\Sqb{12}\Anb{12},\Sab{131}\Sab{232}$
		& 1	& 4,6
\\
&&&\amp+000
	& $\Sqb{12}\Sab{132}$
		& 4	& 6
\\
&&&\amp++00
	& $\Sqb{12}^2$
		& 2	& 6
\\
&&&\amp+-00
	& $\Sab{132}^2 \to \Sab{132}^2-\Asb{132}^2$
		& $\slashed{2}\to1$
				& 8
\myline
$ffvs$  & $14\to 12$    & 12
&\amp++00
	& $\Sqb{12}\{\Sab{313},\Sab{323}\}$
		& 2	& 6
\\
&&&\amp+-00
	& $\Sqb{13}\Anb{23}$
		& 2	& 5
\\
&&&\amp+++0
	& $\Sqb{13}\Sqb{23}$
		& 2	& 6
\\
&&&\amp++-0
	& $\Sqb{12}\Anb{3123} \to\:$\o
		& $\slashed{2}\to0$
				& 8 
\\
&&&\amp+-+0
	& $\Sqb{13}\Sab{312}$
		& 4	& 7
\myline
$ffff$  & 18 & 16
&\amp++++
	& $\Sqb{12}\Sqb{34},\Sqb{13}\Sqb{24}$
			& 2	& 6
\\
&&&\amp++--
	& $\Sqb{12}\Anb{34}$
		& 6	& 6
\\
&&&\amp+++-
	& $\Sqb{12}\Sab{324}$
		& 8	& 7
\myline
$vvvs$  & $35\to27$ & 27
&\amp0000
	& $	\Sqb{12}\Sab{343}\Anb{12},
		\Sqb{13}\Sab{242}\Anb{13},
		\Sqb{23}\Sab{141}\Anb{23}$
			& 1	& 5
\\
&&&\amp+000
	& $\Sqb{12}\Sqb{13}\Anb{23}$
		& 6	& 5
\\
&&&\amp++00
	& $\Sqb{12}^2 \big\{\Sab{313},\Sab{323}\big\}$
		& 6	& 7
\\
&&&\amp+-00
	& $\Sqb{13}\Sab{132}\Anb{23}$
		& $6\to 4$
			& 7
\\
&&&\amp+++0
	& $\Sqb{12}\Sqb{13}\Sqb{23}$
		& 2	& 7
\\
&&&\amp++-0
	& $\Sqb{12}^2\Anb{3123} \to\:$\o
		& $\slashed{6}\to0$
			& 9 
\myline
$vvff$  & $46\to 38$   & 36
&\amp00++
	& $\Anb{12}\times\{\Sqb{12}\Sqb{34},\Sqb{13}\Sqb{24}\}$
		& 2	& 5
\\
&&&\amp00+-
	& $\Anb{14}\Asb{231}\Sqb{23}, \Anb{24}\Asb{132}\Sqb{13}$
		& 2	& 6
\\
&&&\amp0-++
	& $\Anb{12}\Sqb{34}\Asb{241} \to \Anb{12}\Sqb{34}(\Asb{241}/m_1- \Asb{142}/m_2)$
		& $\slashed{4}\to2$
				& 7
\\
&&&\amp0+++
	& $\Asb{132}\times\{\Sqb{12}\Sqb{34},\Sqb{13}\Sqb{24}\}$
		& 4	& 7
\\
&&&\amp0++-
	& $\Anb{14}\Sqb{12}\Sqb{23}$
		& 8	& 6
\\
&&&\amp+++-
	& $\Sqb{12}^2\Sab{314}$
		& 4	& 8
\\
&&&\amp++++
	& $\Sqb{12}\times\{\Sqb{12}\Sqb{34},\Sqb{13}\Sqb{24}\}$
			& 2	& 7
\\
&&&\amp-+++
	& $\Anb{1231}\Sqb{23}\Sqb{24}\to\:$\o
			& $\slashed{4}\to0$
				& 9 
\\
&&&\amp++--
	& $\Sqb{12}^2\Anb{34}$
		& 2	& 7
\\
&&&\amp+--+
	& $\Sqb{14}\Sab{132} \Anb{23} \to \Sqb{14}\Sab{132} \Anb{23}-\Sqb{24}\Sab{231} \Anb{13}$
		& $\slashed{4}\to2$
				& 8
\myline
$vvvv$ & $116\to85$ & 81
&\amp0000
	& $\{\Sqb{12}\Sqb{34},\Sqb{13}\Sqb{24}\}\times
	   \{\Anb{12}\Anb{34},\Anb{13}\Anb{24}\}$
		& 1	& 4
\\
&&&\amp+000
	& $\{\Sqb{12}\Sqb{34} , \Sqb{13}\Sqb{24}\}
	\times \Sab{142} \Anb{34} \to \cdots$
		& $\slashed{8}\to6$
				& 6
\\
&&&\amp++00
	& $\{\Sqb{12}\Sqb{34},\Sqb{13}\Sqb{24}\}
	\times\Sqb{12}\Anb{34}$
		& 12	& 6
\\
&&&\amp+-00
	& $\Sqb{13}\Sqb{14}\Anb{23}\Anb{24}$
		& 12	& 6
\\
&&&\amp+++0
	& $\{\Sqb{12}\Sqb{34},\Sqb{13}\Sqb{24}\}
	\times \Sqb{23} \Sab{134}$
		& 8	& 8
\\
&&&\amp++-0
	& $\Sqb{12}^2 \Anb{34} \Asb{324} \to \Sqb{12}^2 \Anb{34} (\Asb{324} / m_4 - \Asb{423} / m_3) \to\!\cdots$
		& $\cancel{24}\!\to\!\cancel{12}\!\to\!5$
				& 8
\\
&&&\amp++++
	& $\Sqb{12}^2\Sqb{34}^2,\Sqb{12}\Sqb{13}\Sqb{24}\Sqb{34},\Sqb{13}^2\Sqb{24}^2$
			& 2	& 8
\\
&&&\amp+++-
	& $\Sqb{12}\Sqb{13}\Sqb{23}\Anb{4124} \to$ \o
			& $\slashed{8}\to0$
				& 10
\\
&&&\amp++--
	& $\Sqb{12}^2\Anb{34}^2$
		& 6	& 8
\myline
\end{tabular*}
\caption{Spinor structures forming bases of stripped contact terms (SCTs) for four-point scalar, fermion and vector ($s,f,v$) amplitudes.
Arrows keep track of the two reduction steps towards bases.
Helicity categories are distinguished by unbolding spinor structures to obtain massless ones.
For brevity, particle permutations and parity flips are omitted and collectively counted by $n_\text{perm}$.
Accounting for these, the dimension of the bases are provided under $n_\text{SCT}$.
The dimension of spinor-structure bases $n_\text{s}$ are also indicated for reference.
The minimal operator dimension contributing in each case is indicated as $\min\{d_\text{op}\}$.
Most factors of $1/m_i$ in zero-helicity vector categories are left implicit.
}
\label{tab:4pt-massive}
\end{table}

\subsubsection{Stripped-contact-term (SCT) bases}

To construct SCT bases, one can proceed in analogy to the purely massive construction of the spinor-structure bases described above, but eliminate spinor structures only if they are redundant in the SCT sense.
Here too, the reduction can be greatly simplified by using the massless limit.
As in the massless case, the reduction of an over-complete set of spinor structures to an SCT basis is conceptually performed by eliminating, among all possible structures, only those which can be expressed as linear combinations of others with prefactors involving no negative powers of Lorentz invariants.
Therefore, unlike in the construction of spinor-structure bases, one needs to explicitly examine the relations between spinor structures.
A first step of reduction can conveniently be performed by relying mostly on the massless limit, treating one helicity category at a time.
A second step of reduction is possible in some cases, and requires knowledge of the mass-suppressed terms in spinor relations.
In particular, as we will see, inverse powers of the masses can appear when eliminating redundant structures at this stage.
Our results for the SCT bases are summarized in \autoref{tab:4pt-massive}, which lists the structures obtained following the first and second reductions.

\paragraph{Massless and massive spinor relations}
Remarkably, all massive relations between four-point spinor structures have non-trivial massless limits.\footnote{This is different from the three-point case where \autoref{eq:rel_vvv} applies due to the singular three-point kinematics.}
A relation in which each term contains explicit positive powers of the masses could be employed to eliminate the representative of a helicity category in the spinor-basis construction.
The counting of \autoref{eq:n_4pt} would then be invalidated.
Thus, any relation between massive spinors can be obtained by starting with its massless counterpart, and finding the ${\cal O}(m)$ corrections.
We can characterize the relation according to the helicity category of its massless limit.
The mass-suppressed terms involve spinor structures of different helicity categories.
We sometimes refer to the process of finding the ${\cal O}(m)$ corrections to a relation as \emph{mass-completing}.

\paragraph{First reduction}
A first round of reductions can be performed based on the massless limit.
If two massive structures coincide in this limit, one of them can be dropped.
To see this, note that such structures must be in the same helicity category. 
${\cal O}(m)$ corrections to the massless relation necessarily involve spinor-structures of lower dimensions. 
Thus, one of the original structures can be eliminated in favour of the second, plus spinor structures of lower dimension.
As an  example, consider the $\Asb{123}\Sqb{24}$ and $\Asb{124}\Sqb{23}$ structures in the \amp-+++ helicity category of the $ffff$ amplitude.
They both unbold to $\anb{12}\sqb{23}\sqb{24}$ and are equivalent up to a mass correction: $\Asb{123}\Sqb{24} = \Asb{124}\Sqb{23} + m_2 \Anb{12} \Sqb{34}$.
One of them can thus dropped in the SCT basis construction.

This first reduction can be systematized by considering the massless limit of the $P$-matrix.
The relations associated with the eigenvectors of vanishing eigenvalues become relatively simple in this limit, so one can easily identify redundant spinor structures.
Here, these are given by linear combinations of other structures with prefactors containing only non-negative powers of the invariants.
They obviously involve only spinor structures of the same helicity category.
Thus, structures that are redundant in the massless limit can also be eliminated in the massive case, up to mass corrections involving structures of lower dimensions.
The set of independent spinor structures after this first reduction therefore corresponds almost exactly to the massless SCT.
Note that only the massless limits are used at this stage.
The mass corrections to these relations are not needed at all.

One caveat to the above discussion involves massive spinor structures which vanish when unbolded.
These must be examined with special care.
We have found only one example in which the discussion above does not hold, in the \amp0000 helicity category of the $vvvv$ amplitude.
It involves the fully antisymmetric contraction of four polarization vectors  $\epsilon(\bs{\varepsilon_1},\bs{\varepsilon_2},\bs{\varepsilon_3},\bs{\varepsilon_4})$ that unbolds to $\epsilon(p_1,p_2,p_3,p_4)$ which vanishes in a four-point amplitude.
Here, one obtains (using \autoref{eq:rel_ffff} in the second step):
\begin{equation}
\begin{aligned}
\tdp12 & \; i\,\epsilon(\bs{\varepsilon_1},\bs{\varepsilon_2},\bs{\varepsilon_3},\bs{\varepsilon_4}) \:{m_1 m_2 m_3 m_4}
\\
& =\tdp12
	\,\{\,	\Sqb{12}\Sqb{34} \Anb{13}\Anb{24}
	-	\Anb{12}\Anb{34} \Sqb{13}\Sqb{24}
	\,\}
\\
& =
	\begin{array}[t]{*{11}{@{\,}l}}
	+&\Anb{12}\Anb{34}
		&\{\,
		 m_1 &\Asb{123} \Sqb{24}
		&+m_2 &\Asb{213} \Sqb{14}
		&-m_3 &\Asb{324} \Sqb{12}
		&-m_4 &\Asb{423} \Sqb{12}
	\,\}
	\\
	-&\Sqb{12}\Sqb{34}
		&\{\,
		 m_1 &\Sab{123} \Anb{24}
		&+m_2 &\Sab{213} \Anb{14}
		&-m_3 &\Sab{324} \Anb{12}
		&-m_4 &\Sab{423} \Anb{12}
	\}
	\,.
	\end{array}
\end{aligned}
\label{eq:epsilon_0000}
\end{equation}
Thus, $\sqb{12}\sqb{34}\anb{13}\anb{24} = \anb{12}\anb{34}\sqb{13}\sqb{24}$ in the massless limit, but $\Sqb{12}\Sqb{34} \Anb{13}\Anb{24}$ and $\Anb{12}\Anb{34} \Sqb{13}\Sqb{24}$ only become related higher up in the effective-field-theory expansion, once multiplied by an additional power of Mandelstam invariant.
Moreover, since the relation
\begin{multline}
\Sqb{12}\Sqb{34}\Anb{13}\Anb{24} + \Anb{12}\Anb{34}\Sqb{13}\Sqb{24}
\\	= \Sqb{12}\Anb{12}\Sqb{34}\Anb{34}
	- \Sqb{14}\Anb{14}\Sqb{23}\Anb{23}
	+ \Sqb{13}\Anb{13}\Sqb{24}\Anb{24}
\label{eq:peven_0000}
\end{multline}
holds in both the massive and massless cases, four of the five terms it involves are independent in the massive case but only three remain in the massless limit.

\paragraph{Second reduction}

The first reduction described above is adequate for the treatment of spinor structures of maximal helicity categories.
Further reductions are possible however for particles of spin $\geq1$, which feature helicity categories smaller than their spins, or \emph{non-maximal} ($|h_i|<s_i$).
For the amplitudes we consider here, this only occurs for vectors within the zero-helicity category.
To obtain a sensible normalization, the relevant spinor structures are assigned factors of inverse masses, such as 
\beq\label{eq:mass_norm}
\bs i\rangle[\bs i\;/m_i
\eeq
for an external vector $i$.
This combination indeed appears in the massive polarization vectors, $\bs{\varepsilon}_{\bs i} \equiv \sqrt{2}\; {\bs i\rangle [\bs i}\:/{m_i}$.
Including this $1/m_i$ factor is therefore needed to correctly map the spinor-structure dimension to that of the operator generating it.
We can also motivate this normalization directly at the level of on-shell spinor amplitudes.
Longitudinal-polarization amplitudes coincide with scalar amplitudes in the high energy (or massless) limit of Higgsed theories.
These are the Goldstone amplitudes, which scale as $p_i/m_i$ and therefore feature a momentum insertion $i\rangle[ i/m_i$.
Covariantizing this with respect to the massive little group yields $\bs i\rangle[ \bs i/m_i$.
Amplitudes involving such combinations may be badly behaved at high energies.
Indeed, they only arise in isolation in EFTs, like low-energy QCD, with cutoffs not parametrically separated from vector masses.
On the other hand, full amplitudes arising from Higgsed theories like the standard-model EFT can only have energy-growing behaviours matching the highest $1/\Lambda$ power they involve.

In the construction of EFT bases, redundant structures (or operators in the Lagrangian formulation) can be eliminated in favour of lower-dimension ones.
Thus, the inverse mass normalization of \autoref{eq:mass_norm} has important implications from the point of view of constructing the SCT basis, since it affects, and specifically \emph{lowers}, the dimension associated with spinor structures containing ${\bs i}\rangle[ {\bs i}$ combinations.
As a result, additional structures can be eliminated compared to the naive first reduction which relies solely on the massless limit.

After the first step of reduction, a limited number of spinor structures remain in each helicity category.
The second step of reduction generically requires the full massive relations between spinor structures of different helicity categories.
One starts from a massless relation between structures in a given helicity category and mass-completes it.
Though tedious, the latter is a well defined problem.
Fortunately, only a limited number of relations are relevant in this second reduction.
Notably, relations of maximal helicity cannot be used to eliminate any structures.%
\footnote{Recall that the helicity category of the relation is defined by its massless limit.
Starting with all masses in the numerator, the structures surviving in the massless limit determine the helicity category of the relation.}
They do not admit negative powers of the masses and therefore only become relevant at higher dimensions, in the construction of contact terms, where linear combinations involving Lorentz invariants appear.
Only relations involving non-maximal helicity categories, like vectors of zero-helicity categories, thus need to be considered.
Still, they cannot possibly allow for any further reduction if they arise at a dimension higher than that of all of the structures remaining after the first step of reduction.
Eventually, only a limited subset of massless relations therefore needs to be mass-completed.

Let us illustrate this discussion with a few examples.
The first step of reduction left two spinor structures in the \amp++++ helicity category of the $ffff$ amplitude: $\Sqb{12}\Sqb{34}$ and $\Sqb{13}\Sqb{24}$ (see \autoref{tab:4pt-massive}).
The massless relation between their unbolded counterparts, $\tdp24\; \sqb{12} \sqb{34} +\tdp12\; \sqb{13} \sqb{24}=0$, is analogous to \autoref{eq:rel_sij}.
Following the prescription of \autoref{eq:op_dim_massless}, it arises from operators of dimension eight.
No inverse masses are required since the helicity categories of all particles are maximal: $|h_i|=s_i$.
As $ffff$ spinor structures remaining after the first step of reduction have dimension seven at most (see last column of \autoref{tab:4pt-massive}), the dimension-eight relation between $\sqb{12}\sqb{34}$ and $\sqb{13}\sqb{24}$ cannot possibly be used to eliminate any of them.
For the sake of illustration, let us nevertheless examine its mass-completion:
\begin{equation}
\begin{aligned}
0=&\:
  (\tdp24 +m_2^2)\;	\Sqb{12} \Sqb{34}
+ \tdp12\;	\Sqb{13} \Sqb{24}
\\&
+m_1 \Asb{123} \Sqb{24}
+m_2 \Asb{213} \Sqb{14}
-m_3 \Asb{324} \Sqb{12}
-m_4 \Asb{423} \Sqb{12}
\,.
\end{aligned}
\label{eq:rel_ffff}
\end{equation}
As expected, none of the spinor structures it involves can be isolated without introducing negative powers of the Lorentz invariants, or inverse masses not associated with a $\bs i\rangle[\bs i$ combination of spinors.
Insisting on introducing such a negative mass power, one could for instance attempt to eliminate $\Asb{123} \Sqb{24}$ in favour of the other structures appearing in \autoref{eq:rel_ffff}.
This would however correspond to trading a dimension-seven operator for a linear combination involving notably dimension-eight ones.
It is formally possible since, in the presence of dimensionful couplings or masses, equations of motion relate operators of different dimensions.
From the EFT point of view, it is however preferable to exploit redundancies to remove the higher-dimensional operators instead of the lower-dimensional ones.
Following this guideline, we do not allow for inverse masses not arising together with $\bs i\rangle[\bs i$ combinations of spinors.
The relation in \autoref{eq:rel_ffff} therefore only becomes relevant when forming contact terms by appending powers of the Lorentz invariants to SCTs.
At dimension eight, the $\tdp24\; \Sqb{12} \Sqb{34} + \tdp12\; \Sqb{13} \Sqb{24}$ combination is indeed seen to be redundant.

In practice, the second step of reduction towards a SCT basis does not allow to eliminate any spinor structure in amplitudes involving only fermions and scalars.
In these cases, massive SCT bases can directly be obtained by bolding massless ones (collecting for instance the relevant helicity amplitudes in \autoref{tab:4pt-massless-sm}).
As a second example, let us therefore consider the $ffvs$ amplitude involving a vector.
Two $\Sqb{12}\Sab{313}/m_3$ and $\Sqb{12}\Sab{323}/m_3$ spinor structures remain in the \amp++00 helicity category after the first step of reduction.
The presence of a $\bs 3\rangle[\bs 3$ combination of spinors calls for the inverse factor of $m_3$.
Accounting for this, the massless 
$\sqb{12}\sab{313}\tdp23 = \sqb{12}\sab{323}\tdp13$ relation between their unbolded counterparts can be mass-completed to an equality of dimension eight.
Unlike in the previous example, dimension-eight $ffvs$ spinor structures did survive the first reduction in the \amp++-0 and \amp--+0 helicity categories.
The mass-completed relation could possibly allow to eliminate some of these.
Indeed, one finds
\begin{equation}
\begin{aligned}
\Sqb{12}\Anb{3123} &=
	(\Sqb{12} \Sab{313} \tdp23 - \Sqb{12}\Sab{323}\tdp13)/m_3
	\\&
	-\tdp12 \Sqb{13}\Sqb{23}
	-m_1\Sab{321}\Sqb{23}
	-m_2\Sab{312}\Sqb{13}
\,,
\end{aligned}
\label{eq:rel_ffvs}
\end{equation}
which allows to eliminate the $\Sqb{12}\Anb{3123}$ spinor structure~\cite{Durieux:2019eor}.
The parity-flipped analogue of the above equality also allows to eliminate the \amp--+0 structure.

\paragraph{Case-by-case discussion}

In rest of this section, we detail the reduction towards SCT bases for each spin content in turn.
Our results are collected in \autoref{tab:4pt-massive}.
Arrows keep track of the two-step reduction process.
For brevity, we do not explicitly list structures obtained by flipping square and angle brackets (parity), or by permutations of external particles.
These are collectively counted by $n_\text{perm}$.
The number of independent SCTs (i.e.\ the dimension of SCT bases) is provided under $n_\text{SCT}$.
Where it matches the dimension of the spinor-structure basis $n_\text{s}$, the SCT basis can actually be used as spinor-structure basis.
Note however that a non-trivial reduction may still be needed in these case.
We leave implicit the factors of $1/m_i$ required in the presence of $\bs i\rangle[\bs i$ spinor combinations to determine the dimension $d_\text{op}$ of the operator generating each structure.

\subparagraph{\texorpdfstring{$ssss$}{ssss}}
One single spinor structure ---a constant--- appears in the $ssss$ massive SCT basis.

\subparagraph{\texorpdfstring{$vsss$}{vsss}}
The parity-even combination of \amp\pm000 spinor structures is reducible through the mass-completion of the $\tdp13\sab{121} = \tdp12\sab{131}$ equality:
\begin{equation}
\Sqb{1231}+\Anb{1231} =	(  \tdp13 \Sab{121} - \tdp12 \Sab{131} )/m_1 \,.
\label{eq:rel_vsss}
\end{equation}
One can therefore retain only one of the two or, more symmetrically, employ the parity-odd combination:
\begin{equation}
\Sqb{1231}-\Anb{1231} \propto i\:\epsilon(\bs{\varepsilon_1}, p_1, p_2, p_3)
\,.
\label{eq:rel_vsss_podd}
\end{equation}
There thus remain $3$ irreducible spinor structures in the SCT basis.

\subparagraph{\texorpdfstring{$ffss$}{ffss}}
The massive SCT basis is simply obtained by bolding the massless one and contains $4$ elements.

\subparagraph{\texorpdfstring{$vvss$}{vvss}}
The parity even combination of \amp\pm\mp00 spinor structures can be reduced thanks to the mass-completion of the $\sqb{12}\anb{12}\tdp13\tdp23 + \sab{131}\sab{232}\tdp12=0$ equality:
\begin{equation} \label{eqn:vvss_masscompleted}
\begin{aligned}
	\Sab{132}^2
+	\Sab{231}^2
= &
+ \Sqb{12} \Anb{12} \tdp13 \tdp23/m_1m_2
\\&+ \Sab{131} \Sab{232} \tdp12/m_1m_2
\\&
+ (\Sqb{12} \Sab{132}
+ \Anb{12} \Sab{231}) \tdp23 /m_2
\\&- (\Sqb{12} \Sab{231}
+ \Anb{12} \Sab{132}) \tdp13 /m_1
\\&
- (\Sqb{12}^2
+ \Anb{12}^2 ) m_3^2
\,.
\end{aligned}
\end{equation}
The SCT basis therefore eventually contains $9$ massive spinor structures.

\subparagraph{\texorpdfstring{$ffvs$}{ffvs}}
A  spinor-structure basis for the $ffvs$ amplitude was obtained in ref.~\cite{Durieux:2019eor} and also forms a SCT basis.
As discussed above, the \amp\pm\pm\mp0 structures are reducible thanks to the mass-completion of the $\sqb{12}\sab{131}\tdp23=\sqb{12}\sab{323}\tdp13$ equality in \autoref{eq:rel_ffvs}.
One therefore remains with a basis of $12$ spinor structures.

\subparagraph{\texorpdfstring{$ffff$}{ffff}}
As discussed already, no further reduction is possible among the $18$ spinor structures relevant for the $ffff$ amplitude.
Once dressing SCTs with Lorentz invariants to form contact terms, \autoref{eq:rel_ffff} and its parity conjugate make the $ \tdp24\; \Sqb{12} \Sqb{34} + \tdp12\; \Sqb{13} \Sqb{24}$ combination and its parity conjugate redundant at dimension eight.

\subparagraph{\texorpdfstring{$vvvs$}{vvvs}}
The five simplest structures in the \amp0000 helicity category are related via,  
\begin{align}
 \Anb{12}\Asb{341}\Sqb{23}
+\Sqb{12}\Sab{341}\Anb{23}
&=-\Sqb{12}\Sab{343}\Anb{12}
+\Sqb{13}\Sab{242}\Anb{13}
-\Sqb{23}\Sab{141}\Anb{23}
\,,\nonumber\\
\label{eq:vvvsidentity2}
 \Anb{12}\Asb{341}\Sqb{23}
-\Sqb{12}\Sab{341}\Anb{23}
&= m_1(	\Anb{12}\Anb{13}\Sqb{23}
-	\Sqb{12}\Sqb{13}\Anb{23} )\\\nonumber
&+ m_2(	\Anb{12}\Sqb{13}\Anb{23}
-	\Sqb{12}\Anb{13}\Sqb{23} )\\\nonumber
&+ m_3(	\Sqb{12}\Anb{13}\Anb{23}
-	\Anb{12}\Sqb{13}\Sqb{23} )
\,.
\end{align}
As a result, there are only three independent spinor structures in this category.
These relations are most  easily derived by expressing the left-hand sides 
of \autoref{eq:vvvsidentity2} as traces over polarization vectors and momenta, e.g.\ $\Anb{12}\Asb{341}\Sqb{23} \propto \Tr\{\bar{\bs\varepsilon}_{\bs1} \bs{\varepsilon_2} \bar{\bs\varepsilon}_{\bs 3} p_4\}=4(\bs{\varepsilon_1}\cdot\bs{\varepsilon_2})(\bs{\varepsilon_3}\cdot p_4) - 4(\bs{\varepsilon_1}\cdot\bs{\varepsilon_3})(\bs{\varepsilon_2}\cdot p_4) + 4(\bs{\varepsilon_1}\cdot p_4)(\bs{\varepsilon_2}\cdot\bs{\varepsilon_3}) + 2i\epsilon(\bs{\varepsilon_1},\bs{\varepsilon_2},\bs{\varepsilon_3},p_4)$.
Unlike in the similar \amp0000 category of the $vvvv$ amplitude, the number of irreducible massive spinor structures eventually matches that of massless structures.
Multiplying \autoref{eq:rel_ffvs} by $\Sqb{12}$, one can eliminate $\Sqb{12}^2\Anb{3123}$ in the \amp++-0 category.
Two combinations of \amp+-00 structures can moreover be removed\footnote{We thank S.\ Chang, M.\ Chen, D.\ Liu, and M.\ Luty for pointing this out to us.} thanks to the equality
\begin{equation}
\begin{aligned}
	2\,&  (\Sqb{13} \Anb{23} \Asb{231}
	+    \Sqb{23} \Anb{13} \Asb{132} )/m_3
	-2\,  (\Sqb{13} \Anb{12} \Asb{213}
	+    \Sqb{12} \Anb{13} \Asb{312} )/m_1
	\\
	&=  (\Sqb{12} \Anb{13} \Anb{23}
	+    \Sqb{13} \Sqb{23} \Anb{12} ) (\tdp12 + \tdp23) / m_1 m_2 \\
	&-  (\Sqb{12} \Sqb{23} \Anb{13}
	+    \Sqb{13} \Anb{12} \Anb{23} ) (\tdp12 + 2 \tdp13 + \tdp23) / m_1 m_3 \\
	&+  (\Sqb{12} \Sqb{13} \Anb{23}
	+    \Sqb{23} \Anb{12} \Anb{13} ) (\tdp12 + \tdp23) / m_2 m_3 \\
	&+\{(\Sqb{12} \Anb{12} \Asb{343}
	-    \Sqb{23} \Anb{23} \Asb{141})  (\tdp12 + \tdp23)
	-    \Sqb{13} \Anb{13} \Asb{242}   (\tdp12 - \tdp23)
	\}/ m_1 m_2 m_3
\end{aligned}
\end{equation}
and, for instance, its $1\leftrightarrow 2$ permutation.
The SCT basis enventually contains 27 elements, just as the spinor-structure basis.

\subparagraph{\texorpdfstring{$vvff$}{vvff}}
Multiplying~\autoref{eq:rel_ffff} by $\Sqb{12}$ relates the two terms in the \amp++++ category.
The redundant combination $\Sqb{12}^2\Sqb{34}\tdp24 + \Sqb{12}\Sqb{13}\Sqb{24}\tdp12$ and its parity conjugate remain to be eliminated at dimension nine.
Multiplying~\autoref{eq:rel_ffff} by $\Anb{12}/m_1m_2$, the sum of \amp0-++ and \amp-0++ structures can be expressed as,
\begin{equation}
\begin{aligned}
\Anb{12}\Asb{213}&\Sqb{14}/m_1 + \Anb{12}\Asb{123}\Sqb{24}/m_2 
\\&= \Anb{12}\Asb{324}\Sqb{12} m_3/m_1m_2 + \Anb{12}\Asb{423}\Sqb{12} m_4/m_1m_2
\\&- \Anb{12}\Sqb{12}\Sqb{34}(\tdp24 + m_2^2)/m_1m_2 - \Anb{12}\Sqb{13}\Sqb{24}\tdp12/m_1m_2
\,.
\end{aligned}
\end{equation}
One can thus retain only their difference.
This reduces the number of independent particle permutations and parity flips to $2$ instead of $4$ in the \amp0-++ category.
Multiplying~\autoref{eq:rel_ffff} by $\Asb{132}/m_1$ or $\Sab{132}/m_2$ can be used to express the \amp-+++ and \amp+-++ structures in terms of others.
A similar procedure also eliminates their parity conjugates.

The mass-completion of the relation between the two \amp00+- spinor structures $0=\sqb{13} \anb{24} \asb{132} \tdp14 + \sqb{23} \anb{14} \asb{231} \tdp13$ allows us to express the sum of \amp+-+- and \amp-++- structures as,
\begin{equation}
\begin{aligned}
\Sqb{13}& \Anb{24} \Asb{231} + \Sqb{23} \Anb{14} \Asb{132} =	\\
&	- (\Sqb{13} \Anb{24} \Asb{132} (m_1^2 + \tdp14) + \Sqb{23} \Anb{14} \Asb{231} (m_1^2 + \tdp13)) /m_1m_2	\\
&	+ \Sqb{12} (\Sqb{13} \Anb{24} \tdp23 /m_2 - \Sqb{23} \Anb{14} \tdp13 /m_1)\\
&	+ \Sqb{12} \Anb{13} \Anb{24} (m_1^2 + \tdp14)  m_3/m_1m_2	\\
&	- \Sqb{12} \Anb{24} \Asb{321}  m_3/m_2	+ \Sqb{34} \Anb{12} \Asb{132} m_4 /m_2	\\
&	- \Anb{12} (\Sqb{23} \Anb{14} m_3 + \Sqb{24} \Anb{13} m_4) m_3 /m_2	\\
&	- \Sqb{14} \Sqb{23} \Anb{12} \tdp13  m_4/m_1m_2
\,.
\end{aligned}
\end{equation}
One can therefore keep their difference only.

Ultimately, $38$ spinor structures remain in the SCT basis of the $vvff$ amplitude.
There are thus only two additional elements compared to the spinor-structure basis.

\subparagraph{\texorpdfstring{$vvvv$}{vvvv}}

The SCT reduction of the $vvvv$ amplitude is the most involved.
Mass-completing the \amp+++0 category relation $\sqb{23} \asb{431} (\sqb{12} \sqb{34} \tdp24 + \sqb{13}\sqb{24} \tdp12) =0$, one can eliminate the \amp+++- spinor structure,
\begin{equation}%
\adjustbox{max width=1\textwidth}{\ensuremath{%
\begin{gathered}
\Sqb{12} \Sqb{13} \Sqb{23} \Anb{4234}	=
\\
+ \Sqb{12} \Sqb{14} \Sqb{23} \Anb{34} \tdp24 m_3 /m_4
- \Sqb{13} \Sqb{14} \Sqb{23} \Anb{24} (\tdp34 + m_4^2) m_2 /m_4
- \Sqb{13} \Sqb{23} \Sqb{24} \Anb{14} (\tdp34 + m_4^2) m_1 /m_4
\\
+ \Sqb{12} \Sqb{23} \Sqb{34} \Asb{431} \tdp24 /m_4
+ \Sqb{13} \Sqb{23} \Sqb{24} \Asb{431} (\tdp34 + m_4^2) /m_4
- \Sqb{13} \Sqb{14} \Sqb{23} \Asb{214} m_2
- \Sqb{13} \Sqb{23} \Sqb{24} \Asb{124} m_1
\\
- \Sqb{13} \Sqb{14} \Sqb{23} \Sqb{24} (m_3^2 - m_2^2 - m_1^2)
\,.
\end{gathered}}}
\end{equation}
A similar procedure also applies to structures related by particle permutations and parity flips.

The mass-completion of the $\sqb{12} \anb{34} (\sqb{12} \sqb{34} \tdp24 + \sqb{13} \sqb{24} \tdp12)=0$ equality among \amp++00 amplitudes, which directly derives from \autoref{eq:rel_ffff}, removes the sum of the \amp++-0 and \amp++0- spinor structures:
\begin{equation}
\begin{aligned}
\Sqb{12}^2 &\Anb{34} (\Asb{324} / m_4 + \Asb{423} / m_3) = \\
&+ \Sqb{12} \Anb{34} (\Sqb{12} \Sqb{34} (\tdp24 + m_2^2) + \Sqb{13} \Sqb{24} \tdp12) /m_3m_4\\
&+ \Sqb{12} \Anb{34} (\Sqb{14} \Asb{213} m_2 + \Sqb{24} \Asb{123} m_1) /m_3m_4
\,.
\end{aligned}
\end{equation}

The mass-completion of the $\anb{34} \asb{241} (\sqb{12} \sqb{34} \tdp24 + \sqb{13} \sqb{24} \tdp12) = 0$ equality in the \amp+000 category removes the sum of \amp++-0, \amp+-+0, and \amp+-0+ spinor structures:
\begin{equation}
\adjustbox{max width=1\textwidth}{\ensuremath{%
\begin{aligned}
&	  \Sqb{12}^2 \Anb{34} \Asb{314} /m_4 + \Sqb{13}^2 \Anb{24} \Asb{214} /m_4 + \Sqb{14}^2 \Anb{23} \Asb{213} /m_3	=\\
&	- \Sqb{13} \Sqb{24} \Anb{12} \Anb{34} (m_4^2 + \tdp14) m_1 / m_2m_3m_4	\\
&	+ \Sqb{14} (\Sqb{12} \Sqb{23} \Anb{34} (m_4^2 - m_3^2) + \Sqb{13} \Sqb{23} \Anb{24} m_2 m_3 + \Sqb{13} \Sqb{24} \Anb{23} m_2 m_4) /m_3m_4	\\
&	+ \Anb{34} (\Sqb{12} \Sqb{34} \Asb{241} (m_2^2 + \tdp24) + \Sqb{13} \Sqb{24} \Asb{241} (m_2^2 + \tdp12) - \Sqb{23} \Sqb{24} \Asb{141} m_1 m_2) /m_2m_3m_4	\\
&	+ \Anb{34} (\Sqb{12} \Anb{24} \Asb{123} m_4 - \Sqb{14} \Anb{12} \Asb{243} m_2 - \Sqb{24} \Anb{12} \Asb{143} m_1 - \Sqb{24} \Anb{12} \Asb{341} m_3) m_1 /m_2m_3m_4	\\
&	- (\Sqb{12} \Sqb{13} \Anb{24} \Anb{34} \tdp23 m_4 - \Sqb{12} \Sqb{14} \Anb{23} \Anb{34} (m_2^2 + \tdp24) m_3 - \Sqb{12} \Sqb{24} \Anb{13} \Anb{34} m_1 m_2 m_3 \\
&   \quad + \Sqb{13} \Sqb{14} \Anb{23} \Anb{24} (m_4^2 + m_3^2	+ \tdp34) m_2 - \Sqb{13} \Sqb{34} \Anb{12} \Anb{24} m_1 m_2 m_3 + \Sqb{14} \Sqb{34} \Anb{12} \Anb{23} m_1 m_2 m_4) /m_2m_3m_4
\,.
\end{aligned}}}
\end{equation}
Four such equalities are obtained by multiplying \autoref{eq:rel_ffff} by
$\Sab{123} \Anb{24}$,
$\Sab{213} \Anb{14}$,\linebreak[4]
$\Sab{324} \Anb{12}$,
$\Sab{423} \Anb{12}$.
The parity conjugates of these give four additional relations.
However, multiplying \autoref{eq:rel_ffff} by the combination 
$
 m_1 \Sab{123} \Anb{24}
+m_2 \Sab{213} \Anb{14}
-m_3 \Sab{324} \Anb{12}
-m_4 \Sab{423} \Anb{12}$,
and subtracting the parity-conjugate leads to a combination that involves only \amp\pm000 structures and no \amp++-0 one.
Thus, only seven combinations of \amp++-0 structures are actually redundant instead of eight.
Five combinations of the \amp++-0, \amp+-+0, \amp+-0+, \amp-++0, \amp-+0+, \amp-0++ structures therefore remain, which are antisymmetrised under the exchange of vectors in $0$ and $-$ helicity categories.

Four combinations of \amp+000 spinor structures can be expressed in terms of those of the \amp0000 category, as can be seen by multiplying \autoref{eq:rel_ffff} and its parity conjugate by $\Anb{12}\Anb{34},\Anb{13}\Anb{24}$ and $\Sqb{12}\Sqb{34},\Sqb{13}\Sqb{24}$, respectively.
One could for instance remove the four \amp000\pm\ structures.

Eventually, one therefore counts a total of $85$ structures in the SCT basis of the $vvvv$ amplitude, compared to the $81$ of the spinor-structure basis.
Once Mandelstam invariants are appended to SCTs to form the contact terms, four redundancies arise at dimension ten in the \amp++++ category.
They are for instance obtained by multiplying \autoref{eq:rel_ffff} and its parity conjugate by $\Sqb{12}\Sqb{34},\Sqb{13}\Sqb{24}$ and $\Anb{12}\Anb{34},\Anb{13}\Anb{24}$, respectively.

\subsubsection{Contact-term bases}

Finally, contact-term bases are obtained by appending positive powers of the Lorentz invariants to the elements of the SCT bases.
Relations that could not be used to remove spinor structures from the SCT bases, such as \autoref{eq:rel_ffff}, can now be used to eliminate certain combinations at higher dimensions.
They are determined from the massless limit, and no additional mass-completion is required at this stage.
As only two Mandelstam invariants are independent in four-point amplitudes, each of the structures in \autoref{tab:4pt-massive} can for instance be multiplied by increasing powers of  $s_{12}$ and $s_{13}$:
\begin{equation}
\tdp12,\:\tdp13,\qquad
\tdp12^2,\:\tdp12\tdp13,\:\tdp13^2,\qquad
\tdp12^3,\:\tdp12^2\tdp13,\:\tdp12\tdp13^2,\:\tdp13^3,\qquad
\text{etc.}
\end{equation}
So far, we have only considered distinguishable external states.
The (anti)symmetrizations required by spin statistics in the presence of identical particles can however become involved.
These are only performed, in the next section, when specializing our results to the particle content of the electroweak sector of the standard model.
Ultimately, flavour and gauge structures should be combined with the Lorentz structures discussed in this paper.

\section{Electroweak particle content}
\label{sec:ew_application}

Our general results can be specialized to the electroweak sector of the standard model which features a single massive scalar $h$, massive fermions $\psi,\psi'$ and vectors $W^\pm,Z$.
We work in the broken electroweak phase, not imposing the SU(2)$_L\times$ U(1)$_Y$ symmetry.
Electric charge conservation is assumed.
The two species of fermions are taken such that the total electric charge of the $\psi^c\psi'W^+$ vanishes.
Thus, they represent $u,d$ or $\nu,e$ pairs of any generation.
Baryon- and lepton-number conservations are assumed too.
We focus primarily on Lorentz structures generated by operators of dimension six at most.
The minimal dimension at which contributions arise in each helicity category is indicated under $\min\{d_\text{op}\}$ in \autoref{tab:4pt-massive}.
Colour and flavour structures are left implicit and could only be non-trivial for four-fermion operators (massless gluons are not considered).
Spin statistics requires amplitudes involving identical particles to be (anti)symmetric under their exchange.
In the case of fermions, when more than a single generation is considered, the flavour structure can always be chosen to achieve an overall antisymmetrization.
Parity-conjugate structures are not explicitly listed.
Permutations among the helicity category assignments of different particles of identical spins are also understood.

\subparagraph{\texorpdfstring{$ssss$}{ssss}}
The only scalar being the Higgs particle $h$, the only four-point amplitude pertaining to the $ssss$ category has $hhhh$ particle content.
At dimension four, a constant amplitude is generated.
At dimension six, the Bose symmetrization reduces the $\tdp{i}{j}+\text{perm.}$ combination to a sum of square masses.
At dimension eight, a non-trivial amplitude of the form $\tdp12\tdp34+\tdp13\tdp24+\tdp14\tdp23$ is for instance manifestly symmetric.

\subparagraph{\texorpdfstring{$vsss$}{vsss}}
The only electroweak particle content possible for massive $vsss$ amplitudes is $Zhhh$.
In the \amp0000 helicity category, the Bose symmetrization for the three scalars does not allow for a $\Sab{121}/m_1+\text{perm.}$ structure that would have corresponded to an operator of dimension five.
The first amplitudes therefore arise at dimension seven.
An example is the manifestly symmetric $\tdp34\Sab{121}/m_1 + \tdp24\Sab{131}/m_1 + \tdp23\Sab{141}/m_1$ combination.
Note that in the \amp+000 category, the $\Sqb{1231}-\Anb{1231}$ structure also vanishes upon symmetrization given its antisymmetry under the exchange of all three scalars.
This can easily be seen when expressing it as $\epsilon(\bs{\varepsilon_1},p_2,p_3,p_4)$ by using momentum conservation to eliminate $p_1$ in \autoref{eq:rel_vsss_podd}.
An antisymmetric combination of Mandelstam invariants like $(\tdp12-\tdp13)(\tdp12-\tdp14)(\tdp13-\tdp14)$ could be used to form a contact term of dimension thirteen.

\subparagraph{\texorpdfstring{$ffss$}{ffss}}
The $\psic\psi hh$ particle content is the only one possible with $ffss$ spin assignment.
At dimension five, the $\Sqb{12}$ contribution is found in the \amp++00 category, without additional factors of Mandelstam invariants.
At dimension six, the Bose symmetrization does not allow for the $\Sab{132}+(3\leftrightarrow4)$ structure that would otherwise arise in the \amp+-00 helicity category.
This structure only starts contributing at dimension eight, in combinations like $\sab{\bs1(\bs3-\bs4)\bs2} \, (\tdp{1}{3}-\tdp{1}{4})$.

\subparagraph{\texorpdfstring{$vvss$}{vvss}}
Both $ZZhh$ and $W^+W^-hh$ massive particle contents match the $vvss$ spin assignment.
At dimension four, the $\Sqb{12}\Anb{12}/m_1m_2$  structure arises in the \amp0000 helicity category.
The standard-model contact term, which unitarizes this amplitude, is of this form.
At dimension six, the $\tdp34\Sqb{12}\Anb{12}/m_1m_2$ contact term is symmetric under the exchange of the two scalars.
It is incidentally also symmetric under the exchange of the two vectors, so that it is present in both $ZZhh$ and $W^+W^-hh$ cases.
Still at dimension six, one finds $\Sqb{12}^2$ and $(\Sab{131}\Sab{232}+\Sab{141}\Sab{242})/m_1m_2$, in the \amp++00 and \amp0000 categories, which are also symmetric under both vector and scalar exchanges.
The $\tdp34 \Sqb{12}\Anb{12}/m_1 m_2$ and $(\Sab{131}\Sab{232}+\Sab{141}\Sab{242})/m_1m_2$ contact terms which arise in the same helicity category are distinct in the massless limit and therefore independent.
The dimension-six $\Sqb{12}\Sab{132}/m_2$ structure does not survive $3\leftrightarrow4$ symmetrization.

\subparagraph{\texorpdfstring{$ffvs$}{ffvs}}
The $\psic\psi Zh$ and $\psic\psi'W^+h$ particle contents pertain to $ffvs$ massive four-point amplitudes.
In the absence of identical particles, no (anti)symmetrization is required by spin statistics.
At dimension five, the $\Sqb{13}\Anb{23}/m_3$ Lorentz structure arises in the \amp+-00 helicity category.
At dimension six, the $\Sqb{12}\Sab{313}/m_3,\Sqb{12}\Sab{323}/m_3$ and $\Sqb{13}\Sqb{23}$ structures appear respectively in \amp++00 and \amp+++0 categories.

\subparagraph{\texorpdfstring{$ffff$}{ffff}}
The flavour and colour structures of four-fermion amplitudes can be somewhat intricate.
At dimension six, independent $\Sqb{12}\Sqb{34},(\Sqb{13}\Sqb{24}+\Sqb{14}\Sqb{23})$ and $\Sqb{12}\Anb{34}$ Lorentz structures span amplitudes in the \amp++++ and \amp++-- helicity categories.
The $\Sqb{12}\Sqb{34}$,~$\Sqb{12}\Anb{34}$ and $(\Sqb{13}\Sqb{24}+\Sqb{14}\Sqb{23})$ structures are respectively antisymmetric and symmetric under the exchanges of both the first and second pairs of fermions.

\subparagraph{\texorpdfstring{$vvvs$}{vvvs}}
The $W^+W^-Zh$ and $ZZZh$ particle contents belong to $vvvs$ amplitudes.
At dimension five, the $\{\Sqb{12}\Sab{343}\Anb{12}$, $\Sqb{13}\Sab{242}\Anb{13}$, $\Sqb{23}\Sab{141}\Anb{23}\}/m_1m_2m_3$ and $\Sqb{12}\Sqb{13}\Anb{23}/m_2m_3$ independent Lorentz structures respectively arise in the \amp0000 and \amp+000 helicity categories.
The Bose symmetrization required among all three vectors in the $ZZZh$ case reduces the first three structures to their sum $(\Sqb{12}\Sab{343}\Anb{12} + \Sqb{13}\Sab{242}\Anb{13} + \Sqb{23}\Sab{141}\Anb{23})/m_1m_2m_3$.
The fourth one cannot be fully symmetrized (as noted already in the discussion of $WWZ$ and $ZZZ$ amplitudes in ref.~\cite{Durieux:2019eor}).
No new contribution arises at dimension six.

\subparagraph{\texorpdfstring{$vvff$}{vvff}}
The three $ZZ\psic\psi$, $W^+W^-\psic\psi$, $W^+Z\psic\psi'$ particle contents appear in massive four-point amplitudes of $vvff$ type.
At dimension five, the two $\Anb{12}\times(\Sqb{12}\Sqb{34},\Sqb{13}\Sqb{24}+\Sqb{14}\Sqb{23})/m_1m_2$ Lorentz structures span the \amp00++ helicity category.
The second drops, at dimension five, after the symmetrization required in the $ZZ\psic\psi$ case.
At dimension six, the two $\{\Anb{14}\Asb{231}\Sqb{23}$, $\Anb{24}\Asb{132}\Sqb{13}\}/m_1m_2$ Lorentz structures arise in the \amp00+- category.
Only their sum survives symmetrization under the two vector exchange.
Still at dimension six, the $\Anb{14}\Sqb{12}\Sqb{23}/m_1$ structure arises in the \amp0++- helicity category and can straightforwardly be symmetrized under the exchange of the two vectors.

\subparagraph{\texorpdfstring{$vvvv$}{vvvv}}
The three $W^+W^+W^-W^-$, $W^+W^-ZZ$, $ZZZZ$ electroweak massive particle contents are of $vvvv$ type.
At dimension four, four independent Lorentz structures appear in the \amp0000 helicity category: $\{\Sqb{12}\Sqb{34},\Sqb{13}\Sqb{24}\}\times \{\Anb{12}\Anb{34},\Anb{13}\Anb{24}\}/m_1m_2m_3m_4$.
Two remain upon symmetrization over the last two vectors: e.g.\ $\Sqb{12}\Sqb{34}\Anb{12}\Anb{34}/m_1m_2m_3m_4$ and $(\Sqb{13}\Sqb{24}\Anb{13}\Anb{24}+\Sqb{14}\Sqb{23}\Anb{14}\Anb{23})/m_1m_2m_3m_4$.
These are incidentally also symmetric under the exchange of the first two vectors and therefore apply to both $W^+W^-ZZ$ and $W^+W^+W^-W^-$ cases.
A full symmetrization, required for $ZZZZ$, leaves only one independent structure: $(\Sqb{12}\Sqb{34}\Anb{12}\Anb{34}+\Sqb{13}\Sqb{24}\Anb{13}\Anb{24}+\Sqb{14}\Sqb{23}\Anb{14}\Anb{23})/m_1m_2m_3m_4$.
At dimension six, additional Lorentz structures are generated in the \amp0000, \amp+000, \amp+-00 and \amp++00 helicity categories.
In the \amp0000 case, one power of Mandelstam invariant can be appended to the dimension-four Lorentz structures just described.
After symmetrization under the first and last two pairs of vectors, one for instance obtains the three $(\tdp12+\tdp34)\{\Sqb{12}\Sqb{34}\Anb{12}\Anb{34},\Sqb{13}\Sqb{24}\Anb{13}\Anb{24}+\Sqb{14}\Sqb{23}\Anb{14}\Anb{23}\}/m_1m_2m_3m_4$ and $(\tdp13-\tdp14-\tdp23+\tdp24)(\Sqb{13}\Sqb{24}\Anb{13}\Anb{24}-\Sqb{14}\Sqb{23}\Anb{14}\Anb{23})/m_1m_2m_3m_4$ structures.
No such fully symmetric \amp0000 structure can be formed at dimension six.
In the \amp++00 case, there is just one Lorentz structure which is symmetric under the exchange of the first and last two vectors: $\Sqb{12}^2\Sqb{34}\Anb{34}/m_3m_4$.
It can easily be symmetrized under other permutations too.
A full symmetrization is straightforward and involves all six \amp++00,\,\amp+0+0,\,\amp+00+,\,\amp0++0,\,\amp0+0+,\,\amp00++ helicity categories.
To treat the \amp+000 case, it is useful to re-express \autoref{eq:rel_ffff} in a more symmetric form:
\begin{equation}
\begin{gathered}
  m_1 [\bs1(\bs3-\bs4)\bs2\rangle \Anb{34}
+ m_2 \langle\bs1(\bs3-\bs4)\bs2] \Anb{34}
+ m_3 \Anb{12} [\bs3(\bs1-\bs2)\bs4\rangle
+ m_4 \Anb{12} \langle\bs3(\bs1-\bs2)\bs4]
\\
-2 m_1 m_3 \Sqb{13} \Anb{24}
-2 m_2 m_4 \Anb{13} \Sqb{24}
-2 m_1 m_4 \Sqb{14} \Anb{23}
-2 m_2 m_3 \Anb{14} \Sqb{23}
\\
+(\tdp{1}{2}+\tdp{1}{3}-\tdp{1}{4} + m_1^2 + m_2^2 + m_3^2 - m_4^2) \Anb{13} \Anb{24} \\
+(\tdp{1}{2}-\tdp{1}{3}+\tdp{1}{4} + m_1^2 + m_2^2 - m_3^2 +                                                                                                                                    m_4^2) \Anb{14} \Anb{23}
=0\,.
\end{gathered}
\end{equation}
Multiplied by $\Sqb{13}\Sqb{24}+\Sqb{14}\Sqb{23}$, the first line involves two combinations of \amp+000,\amp0+00 and \amp00+0,\amp000+ Lorentz structures that are symmetric under the first and last two vector exchanges.
Since their sum can be expressed in terms of structures already included, one can for instance retain their difference: $  (m_1 [\bs1(\bs3-\bs4)\bs2\rangle \Anb{34}
+ m_2 [\bs1(\bs3-\bs4)\bs2\rangle \Anb{34}
- m_3 \Anb{12} [\bs3(\bs1-\bs2)\bs4\rangle
- m_4 \Anb{12} \langle\bs3(\bs1-\bs2)\bs4])(\Sqb{13}\Sqb{24}+\Sqb{14}\Sqb{23})/m_1m_2m_3m_4$.
This Lorentz structure can also be fully symmetrized to apply to the $ZZZZ$ case.
In the last \amp+-00 helicity category, which receives dimension-six contributions, the $\Sqb{13}\Sqb{14}\Anb{23}\Anb{24}/m_3m_4$ structure is symmetric under the exchange of the last two vectors.
This structure can easily be symmetrized under any other permutation.

\section{Conclusions}

We addressed the construction of contact-term bases required to form massive on-shell three- and four-point amplitudes.
This is a crucial step in the application of on-shell methods to the classification and computation of EFT amplitudes.
We derived general  expressions for four-point amplitudes involving massive scalars, spin-$1/2$ fermions and vectors, as well as for all three-point amplitudes involving particles of spins $\leq3$.
Some of the techniques we described also apply to higher-spin and higher-point contact terms.
The massless limit was extensively exploited.
Its most powerful application is the derivation of the independent building blocks of generic amplitudes: spinor-structure bases.
These can be obtained from appropriate sets of massless amplitudes written in terms of massless spinors, by \emph{bolding} the spinors into the massive spinors of ref.~\cite{Arkani-Hamed:2017jhn}.
A similar procedure applies to the derivation of contact terms involving scalars and fermions only.
Contact terms featuring vectors and higher spins are more challenging, and require non-trivial massive-spinor identities.
Here too, the massless limit provides a useful starting point, but finding the mass corrections to the massless identities requires a case-by-case, and often tedious, analysis.
It would be interesting to examine whether a more direct and algorithmic construction could be achieved.
Bypassing the intermediate construction of stripped-contact-term bases would be particularly useful, perhaps by extending the massless harmonics approach of refs.~\cite{Henning:2019enq, Henning:2019mcv} to massive amplitudes.
Streamlining  the (anti)symmetrization required in the presence of identical particles, preferably so this is built-in in the construction of the contact terms, would also be useful.

The results presented here already allow for a wide range of applications.
The ingredients we provide, in the form of independent spinor structures, can be used to derive the most general four-point EFT amplitudes for spin $\leq1$ particles, up to any dimension, simply by appending Lorentz invariants.
These basic spinor structures, which carry all the spin information, are furthermore given by very compact expressions.
Thus, various properties of the amplitudes, such as the leading mass effects in their interference with the standard-model amplitudes, can be easily read-off.

For the electroweak bosons and fermions, we also perform the required (anti)symmetrization of all contact terms up to dimension six, including also few notable higher-dimensional ones.
This results in interesting patterns, with some structures only appearing at very high dimensions.
It would be interesting to see how these patterns are further restricted by the SU(2)$_L\times$U(1)$_Y$ symmetry, and to identify sensitive probes to differentiate between these different EFTs.
Since they furthermore encode the full $v/\Lambda$ expansion, the amplitudes we derived  are interesting starting points for the exploration of general EFT extensions of the standard model.

\section*{Acknowledgments}
C.S.M.\ thanks the Technion Particle Physics group for  hospitality while  this work was initiated.
Research supported in part by the Israel Science Foundation (Grant No.\,751/19). 
The work of G.D.\ is supported in part at the Technion by a fellowship from the Lady Davis Foundation.
The work of T.K.\ is supported by the Japan Society for the Promotion of Science (JSPS) Grant-in-Aid for Early-Career Scientists (Grant No.\,19K14706) and the JSPS Core-to-Core Program (Grant No.\,JPJSCCA20200002). The work of C.S.M.\ is supported by the Alexander von Humboldt Foundation, in the framework of the Sofja Kovalevskaja Award 2016, endowed by the German Federal Ministry of Education and Research and also supported by  the  Cluster  of  Excellence  ``Precision  Physics,  Fundamental Interactions, and Structure of Matter'' (PRISMA$^+$ EXC 2118/1) funded by the German Research Foundation (DFG) within the German Excellence Strategy (Project ID 39083149).

\begin{table}
\adjustbox{max width=\textwidth}{%
\ensuremath{\begin{array}{@{}c@{\;}c@{\;}cccc@{}}
s_1	& s_2	& s_3	& n^\text{3-pt}	& n_\text{rel}	& \text{spinor structures}\\
\input{3pt.tex}
\end{array}}}
\caption{Massive three-point amplitudes for spins up to $3$.
For each combinations of three spins $s_1,s_2,s_3$, the number of independent spinor structures $n^\text{3-pt}$ is indicated together with the spinor structures themselves, in a schematic \emph{factorized} way.
When relations based on \autoref{eq:rel_vvv} apply, their number is indicated under $n_\text{rel}$.
They are obtained by multiplying \autoref{eq:rel_vvv} by each of the $s_1-1,s_2-1,s_3-1$ structures appearing higher up in the table.
}
\label{tab:3pt}
\end{table}

\newcommand{\mytab}[2]{%
\begin{table}[t]\centering
\adjustbox{max height=.45\textheight}{%
\ensuremath{\begin{array}{c@{\quad} *{4}{c@{\,}} @{\quad} c}
\text{min.\ operator dim.}
	& \multicolumn{4}{c@{\quad}}{\text{helicities}}
				&\text{spinor structures}
\\[2mm]\hline\noalign{\vskip2mm}
\input{#1}
\\[2mm]\hline
\end{array}}}
\caption{#2}
\label{tab:#1}
\end{table}}

\begin{table}[t]\centering
\adjustbox{max height=.45\textheight}{%
\ensuremath{\begin{array}{c@{\quad} *{4}{c@{\,}} @{\quad} c}
\text{min.\ operator dim.}
	& \multicolumn{4}{c@{\quad}}{\text{helicities}}
				&\text{spinor structures}
\\[2mm]\hline\noalign{\vskip2mm}
\input{4pt-massless-sm}
\\[2mm]\hline
\end{array}}}
\caption{Spinor structures forming bases of massless four-point stripped contact terms for spins $0,1/2,1$, and obtained by the algorithmic procedure described in ref.~\cite{Durieux:2019siw} and \autoref{sec:4pt-massless}.
Only structures with non-negative total helicity are displayed.
Others are obtained by parity conjugation.
Massless four-point bases for amplitudes involving spin-$3/2$ and $2$ are respectively provided in \autoref{tab:4pt-massless-three-half} and \ref{tab:4pt-massless-two}.
}
\label{tab:4pt-massless-sm}
\end{table}

\begin{table}[t]\centering
\adjustbox{max height=.45\textheight}{%
\ensuremath{\begin{array}{c@{\quad} *{4}{c@{\,}} @{\quad} c}
\text{min.\ operator dim.}
	& \multicolumn{4}{c@{\quad}}{\text{helicities}}
				&\text{spinor structures}
\\[2mm]\hline\noalign{\vskip2mm}
\input{4pt-massless-three-half}
\\[2mm]\hline
\end{array}}}
\caption{Spinor structures forming bases of massless four-point stripped contact terms for particles of spin $3/2$ at most.
At least a particle of spin $3/2$ must be involved.
For spin $1$ at most, see \autoref{tab:4pt-massless-sm}.}
\label{tab:4pt-massless-three-half}
\end{table}

\begin{table}[t]\centering
\adjustbox{max height=.45\textheight}{%
\ensuremath{\begin{array}{c@{\quad} *{4}{c@{\,}} @{\quad} c}
\text{min.\ operator dim.}
	& \multicolumn{4}{c@{\quad}}{\text{helicities}}
				&\text{spinor structures}
\\[2mm]\hline\noalign{\vskip2mm}
\input{4pt-massless-two}
\\[2mm]\hline
\end{array}}}
\caption{Spinor structures forming bases of massless four-point stripped contact terms for particles of spin $2$ at most.
At least a particle of spin $2$ must be involved.
For spin $1$ and spin $3/2$ at most, see \autoref{tab:4pt-massless-sm} and \ref{tab:4pt-massless-three-half}.}
\label{tab:4pt-massless-two}
\end{table}

\bibliographystyle{apsrev4-1_title}
\bibliography{notes}
\end{document}

%% file: 3pt.tex
 0 & 0 & 0 	& 1	&	&\textrm{constant}\\
 0 & 0 & 1 	& 1	&	&[\bs3(\bs1-\bs2)\bs3\rangle\\
 0 & 0 & 2 	& 1	&	&[\bs3(\bs1-\bs2)\bs3\rangle^2\\
 0 & 0 & 3 	& 1	&	&[\bs3(\bs1-\bs2)\bs3\rangle^3\\
 0 &1/2&1/2	& 2	&	&([\bs{23}],\anb{\bs{23}})\\
 0 &1/2&3/2	& 2	&	&[\bs3(\bs1-\bs2)\bs3\rangle\otimes([\bs{23}],\anb{\bs{23}})\\
 0 &1/2&5/2	& 2	&	&[\bs3(\bs1-\bs2)\bs3\rangle^2\otimes([\bs{23}],\anb{\bs{23}})\\
 0 & 1 & 1 	& 3	&	&([\bs{23}]^2,\anb{\bs{23}}[\bs{23}],\anb{\bs{23}}^2)\\
 0 & 1 & 2 	& 3	&	&[\bs3(\bs1-\bs2)\bs3\rangle\otimes([\bs{23}]^2,\anb{\bs{23}}[\bs{23}],\anb{\bs{23}}^2)\\
 0 & 1 & 3 	& 3	&	&[\bs3(\bs1-\bs2)\bs3\rangle^2\otimes([\bs{23}]^2,\anb{\bs{23}}[\bs{23}],\anb{\bs{23}}^2)\\
 0 &3/2&3/2	& 4	&	&([\bs{23}]^3,\anb{\bs{23}}[\bs{23}]^2,\anb{\bs{23}}^2[\bs{23}],\anb{\bs{23}}^3)\\
 0 &3/2&5/2	& 4	&	&[\bs3(\bs1-\bs2)\bs3\rangle\otimes([\bs{23}]^3,\anb{\bs{23}}[\bs{23}]^2,\anb{\bs{23}}^2[\bs{23}],\anb{\bs{23}}^3)\\
 0 & 2 & 2 	& 5	&	&([\bs{23}]^4,\anb{\bs{23}}[\bs{23}]^3,\anb{\bs{23}}^2[\bs{23}]^2,\anb{\bs{23}}^3[\bs{23}],\anb{\bs{23}}^4)\\
 0 & 2 & 3 	& 5	&	&[\bs3(\bs1-\bs2)\bs3\rangle\otimes([\bs{23}]^4,\anb{\bs{23}}[\bs{23}]^3,\anb{\bs{23}}^2[\bs{23}]^2,\anb{\bs{23}}^3[\bs{23}],\anb{\bs{23}}^4)\\
 0 &5/2&5/2	& 6	&	&([\bs{23}]^5,\anb{\bs{23}}[\bs{23}]^4,\anb{\bs{23}}^2[\bs{23}]^3,\anb{\bs{23}}^3[\bs{23}]^2,\anb{\bs{23}}^4[\bs{23}],\anb{\bs{23}}^5)\\
 0 & 3 & 3 	& 7	&	&([\bs{23}]^6,\anb{\bs{23}}[\bs{23}]^5,\anb{\bs{23}}^2[\bs{23}]^4,\anb{\bs{23}}^3[\bs{23}]^3,\anb{\bs{23}}^4[\bs{23}]^2,\anb{\bs{23}}^5[\bs{23}],\anb{\bs{23}}^6)\\
1/2&1/2& 1 	& 4	&	&([\bs{23}],\anb{\bs{23}})\otimes([\bs{13}],\anb{\bs{13}})\\
1/2&1/2& 2 	& 4	&	&[\bs3(\bs1-\bs2)\bs3\rangle\otimes([\bs{23}],\anb{\bs{23}})\otimes([\bs{13}],\anb{\bs{13}})\\
1/2&1/2& 3 	& 4	&	&[\bs3(\bs1-\bs2)\bs3\rangle^2\otimes([\bs{23}],\anb{\bs{23}})\otimes([\bs{13}],\anb{\bs{13}})\\
1/2& 1 &3/2	& 6	&	&([\bs{23}]^2,\anb{\bs{23}}[\bs{23}],\anb{\bs{23}}^2)\otimes([\bs{13}],\anb{\bs{13}})\\
1/2& 1 &5/2	& 6	&	&[\bs3(\bs1-\bs2)\bs3\rangle\otimes([\bs{23}]^2,\anb{\bs{23}}[\bs{23}],\anb{\bs{23}}^2)\otimes([\bs{13}],\anb{\bs{13}})\\
1/2&3/2& 2 	& 8	&	&([\bs{23}]^3,\anb{\bs{23}}[\bs{23}]^2,\anb{\bs{23}}^2[\bs{23}],\anb{\bs{23}}^3)\otimes([\bs{13}],\anb{\bs{13}})\\
1/2&3/2& 3 	& 8	&	&[\bs3(\bs1-\bs2)\bs3\rangle\otimes([\bs{23}]^3,\anb{\bs{23}}[\bs{23}]^2,\anb{\bs{23}}^2[\bs{23}],\anb{\bs{23}}^3)\otimes([\bs{13}],\anb{\bs{13}})\\
1/2& 2 &5/2	&10	&	&([\bs{23}]^4,\anb{\bs{23}}[\bs{23}]^3,\anb{\bs{23}}^2[\bs{23}]^2,\anb{\bs{23}}^3[\bs{23}],\anb{\bs{23}}^4)\otimes([\bs{13}],\anb{\bs{13}})\\
1/2&5/2& 3 	&12	&	&([\bs{23}]^5,\anb{\bs{23}}[\bs{23}]^4,\anb{\bs{23}}^2[\bs{23}]^3,\anb{\bs{23}}^3[\bs{23}]^2,\anb{\bs{23}}^4[\bs{23}],\anb{\bs{23}}^5)\otimes([\bs{13}],\anb{\bs{13}})\\
 1 & 1 & 1 	& 7	&1	&([\bs{12}],\anb{\bs{12}})\otimes([\bs{23}],\anb{\bs{23}})\otimes([\bs{13}],\anb{\bs{13}})\\
 1 & 1 & 2 	& 9	&	&([\bs{23}]^2,\anb{\bs{23}}[\bs{23}],\anb{\bs{23}}^2)\otimes([\bs{13}]^2,\anb{\bs{13}}[\bs{13}],\anb{\bs{13}}^2)\\
 1 & 1 & 3 	& 9	&	&[\bs3(\bs1-\bs2)\bs3\rangle\otimes([\bs{23}]^2,\anb{\bs{23}}[\bs{23}],\anb{\bs{23}}^2)\otimes([\bs{13}]^2,\anb{\bs{13}}[\bs{13}],\anb{\bs{13}}^2)\\
 1 &3/2&3/2	&10	&2	&([\bs{12}],\anb{\bs{12}})\otimes([\bs{23}]^2,\anb{\bs{23}}[\bs{23}],\anb{\bs{23}}^2)\otimes([\bs{13}],\anb{\bs{13}})\\
 1 &3/2&5/2	&12	&	&([\bs{23}]^3,\anb{\bs{23}}[\bs{23}]^2,\anb{\bs{23}}^2[\bs{23}],\anb{\bs{23}}^3)\otimes([\bs{13}]^2,\anb{\bs{13}}[\bs{13}],\anb{\bs{13}}^2)\\
 1 & 2 & 2 	&13	&3	&([\bs{12}],\anb{\bs{12}})\otimes([\bs{23}]^3,\anb{\bs{23}}[\bs{23}]^2,\anb{\bs{23}}^2[\bs{23}],\anb{\bs{23}}^3)\otimes([\bs{13}],\anb{\bs{13}})\\
 1 & 2 & 3 	&15	&	&([\bs{23}]^4,\anb{\bs{23}}[\bs{23}]^3,\anb{\bs{23}}^2[\bs{23}]^2,\anb{\bs{23}}^3[\bs{23}],\anb{\bs{23}}^4)\otimes([\bs{13}]^2,\anb{\bs{13}}[\bs{13}],\anb{\bs{13}}^2)\\
 1 &5/2&5/2	&16	&4	&([\bs{12}],\anb{\bs{12}})\otimes([\bs{23}]^4,\anb{\bs{23}}[\bs{23}]^3,\anb{\bs{23}}^2[\bs{23}]^2,\anb{\bs{23}}^3[\bs{23}],\anb{\bs{23}}^4)\otimes([\bs{13}],\anb{\bs{13}})\\
 1 & 3 & 3 	&19	&5	&([\bs{12}],\anb{\bs{12}})\otimes([\bs{23}]^5,\anb{\bs{23}}[\bs{23}]^4,\anb{\bs{23}}^2[\bs{23}]^3,\anb{\bs{23}}^3[\bs{23}]^2,\anb{\bs{23}}^4[\bs{23}],\anb{\bs{23}}^5)\otimes([\bs{13}],\anb{\bs{13}})\\
3/2&3/2& 2 	&14	&4	&([\bs{12}],\anb{\bs{12}})\otimes([\bs{23}]^2,\anb{\bs{23}}[\bs{23}],\anb{\bs{23}}^2)\otimes([\bs{13}]^2,\anb{\bs{13}}[\bs{13}],\anb{\bs{13}}^2)\\
3/2&3/2& 3 	&16	&	&([\bs{23}]^3,\anb{\bs{23}}[\bs{23}]^2,\anb{\bs{23}}^2[\bs{23}],\anb{\bs{23}}^3)\otimes([\bs{13}]^3,\anb{\bs{13}}[\bs{13}]^2,\anb{\bs{13}}^2[\bs{13}],\anb{\bs{13}}^3)\\
3/2& 2 &5/2	&18	&6	&([\bs{12}],\anb{\bs{12}})\otimes([\bs{23}]^3,\anb{\bs{23}}[\bs{23}]^2,\anb{\bs{23}}^2[\bs{23}],\anb{\bs{23}}^3)\otimes([\bs{13}]^2,\anb{\bs{13}}[\bs{13}],\anb{\bs{13}}^2)\\
3/2&5/2& 3 	&22	&8	&([\bs{12}],\anb{\bs{12}})\otimes([\bs{23}]^4,\anb{\bs{23}}[\bs{23}]^3,\anb{\bs{23}}^2[\bs{23}]^2,\anb{\bs{23}}^3[\bs{23}],\anb{\bs{23}}^4)\otimes([\bs{13}]^2,\anb{\bs{13}}[\bs{13}],\anb{\bs{13}}^2)\\
 2 & 2 & 2 	&19	&8	&([\bs{12}]^2,\anb{\bs{12}}[\bs{12}],\anb{\bs{12}}^2)\otimes([\bs{23}]^2,\anb{\bs{23}}[\bs{23}],\anb{\bs{23}}^2)\otimes([\bs{13}]^2,\anb{\bs{13}}[\bs{13}],\anb{\bs{13}}^2)\\
 2 & 2 & 3 	&23	&9	&([\bs{12}],\anb{\bs{12}})\otimes([\bs{23}]^3,\anb{\bs{23}}[\bs{23}]^2,\anb{\bs{23}}^2[\bs{23}],\anb{\bs{23}}^3)\otimes([\bs{13}]^3,\anb{\bs{13}}[\bs{13}]^2,\anb{\bs{13}}^2[\bs{13}],\anb{\bs{13}}^3)\\
 2 &5/2&5/2	&24	&12	&([\bs{12}]^2,\anb{\bs{12}}[\bs{12}],\anb{\bs{12}}^2)\otimes([\bs{23}]^3,\anb{\bs{23}}[\bs{23}]^2,\anb{\bs{23}}^2[\bs{23}],\anb{\bs{23}}^3)\otimes([\bs{13}]^2,\anb{\bs{13}}[\bs{13}],\anb{\bs{13}}^2)\\
 2 & 3 & 3 	&29	&16	&([\bs{12}]^2,\anb{\bs{12}}[\bs{12}],\anb{\bs{12}}^2)\otimes([\bs{23}]^4,\anb{\bs{23}}[\bs{23}]^3,\anb{\bs{23}}^2[\bs{23}]^2,\anb{\bs{23}}^3[\bs{23}],\anb{\bs{23}}^4)\otimes([\bs{13}]^2,\anb{\bs{13}}[\bs{13}],\anb{\bs{13}}^2)\\
5/2&5/2& 3 	&30	&18	&([\bs{12}]^2,\anb{\bs{12}}[\bs{12}],\anb{\bs{12}}^2)\otimes([\bs{23}]^3,\anb{\bs{23}}[\bs{23}]^2,\anb{\bs{23}}^2[\bs{23}],\anb{\bs{23}}^3)\otimes([\bs{13}]^3,\anb{\bs{13}}[\bs{13}]^2,\anb{\bs{13}}^2[\bs{13}],\anb{\bs{13}}^3)\\
 3 & 3 & 3 	&37	&27	&([\bs{12}]^3,\anb{\bs{12}}[\bs{12}]^2,\anb{\bs{12}}^2[\bs{12}],\anb{\bs{12}}^3)\otimes([\bs{23}]^3,\anb{\bs{23}}[\bs{23}]^2,\anb{\bs{23}}^2[\bs{23}],\anb{\bs{23}}^3)\otimes([\bs{13}]^3,\anb{\bs{13}}[\bs{13}]^2,\anb{\bs{13}}^2[\bs{13}],\anb{\bs{13}}^3)\\

%% file: 4pt-massless-sm.tex
\text{dim-4}
	&  0  &  0  &  0  &  0 	&\text{constant}\\
\text{dim-5}
	& 1/2 & 1/2 &  0  &  0 	&\sqb{12}\\
\text{dim-6}
	&  1  &  1  &  0  &  0 	&\sqb{12}^2\\
	&  1  & 1/2 & 1/2 &  0 	&\sqb{12} \sqb{13}\\
	& 1/2 & 1/2 & 1/2 & 1/2	&\sqb{12} \sqb{34}, \sqb{13} \sqb{24}\\
	& 1/2 & 1/2 &-1/2 &-1/2	&\sqb{12} \anb{34}\\
	& 1/2 &-1/2 &  0  &  0 	&\sqb{13} \anb{23}\\
\text{dim-7}
	&  1  &  1  &  1  &  0 	&\sqb{12} \sqb{13} \sqb{23}\\
	&  1  &  1  & 1/2 & 1/2	&\sqb{12}^2 \sqb{34}, \sqb{12} \sqb{13} \sqb{24}\\
	&  1  &  1  &-1/2 &-1/2	&\sqb{12}^2 \anb{34}\\
	&  1  & 1/2 &-1/2 &  0 	&\sqb{12}^2 \anb{23}\\
	&  1  &  0  &  0  &  0 	&\sqb{12} \sqb{13} \anb{23}\\
	& 1/2 & 1/2 & 1/2 &-1/2	&\sqb{12} \sqb{23} \anb{24}\\
\text{dim-8}
	&  1  &  1  &  1  &  1 	&\sqb{12}^2 \sqb{34}^2, \sqb{12} \sqb{13} \sqb{24} \sqb{34}, \sqb{13}^2 \sqb{24}^2\\
	&  1  &  1  & -1  & -1 	&\sqb{12}^2 \anb{34}^2\\
	&  1  &  1  & 1/2 &-1/2	&\sqb{12}^2 \sqb{23} \anb{24}\\
	&  1  & -1  & 1/2 &-1/2	&\sqb{13}^2 \anb{23} \anb{24}\\
	&  1  & -1  &  0  &  0 	&\sqb{13}^2 \anb{23}^2\\
	&  1  &-1/2 &-1/2 &  0 	&\sqb{12} \sqb{13} \anb{23}^2\\
	& -1  & 1/2 & 1/2 &  0 	&\anb{12} \anb{13} \sqb{23}^2\\
\text{dim-9}
	&  1  &  1  & -1  &  0 	&\sqb{12}^3 \anb{13} \anb{23}\\
	&  1  & -1  & 1/2 & 1/2	&\sqb{13}^2 \anb{23}^2 \sqb{34}\\
\text{dim-10}
	&  1  &  1  &  1  & -1 	&\sqb{12}^2 \sqb{23}^2 \anb{24}^2

%% file: 4pt-massless-three-half.tex
\text{dim-5}
	& 3/2 & 3/2 &  0  &  0 	&\sqb{12}^3\\
\text{dim-6}
	& 3/2 & 3/2 & 3/2 & 3/2	&\sqb{12} \sqb{13}^2 \sqb{24}^2 \sqb{34}, \sqb{12}^3 \sqb{34}^3, \sqb{13}^3 \sqb{24}^3, \sqb{12}^2 \sqb{13} \sqb{24} \sqb{34}^2\\
	& 3/2 & 3/2 & 3/2 & 1/2	&\sqb{12} \sqb{13}^2 \sqb{23} \sqb{24}, \sqb{12}^2 \sqb{13} \sqb{23} \sqb{34}\\
	& 3/2 & 3/2 &-3/2 &-3/2	&\sqb{12}^3 \anb{34}^3\\
	& 3/2 & 3/2 &  1  &  0 	&\sqb{12}^2 \sqb{13} \sqb{23}\\
	& 3/2 & 3/2 & 1/2 & 1/2	&\sqb{12}^2 \sqb{13} \sqb{24}, \sqb{12}^3 \sqb{34}\\
	& 3/2 & 3/2 &-1/2 &-1/2	&\sqb{12}^3 \anb{34}\\
	& 3/2 &  1  & 1/2 &  0 	&\sqb{12}^2 \sqb{13}\\
	& 3/2 & 1/2 & 1/2 & 1/2	&\sqb{12} \sqb{13} \sqb{14}\\
\text{dim-7}
	& 3/2 & 3/2 & 3/2 &-1/2	&\sqb{12}^2 \sqb{13} \sqb{23}^2 \anb{24}\\
	& 3/2 & 3/2 &  1  &  1 	&\sqb{12}^3 \sqb{34}^2, \sqb{12} \sqb{13}^2 \sqb{24}^2, \sqb{12}^2 \sqb{13} \sqb{24} \sqb{34}\\
	& 3/2 & 3/2 & -1  & -1 	&\sqb{12}^3 \anb{34}^2\\
	& 3/2 & 3/2 & 1/2 &-1/2	&\sqb{12}^3 \sqb{23} \anb{24}\\
	& 3/2 &  1  &  1  & 1/2	&\sqb{12} \sqb{13}^2 \sqb{24}, \sqb{12}^2 \sqb{13} \sqb{34}\\
	& 3/2 &  1  &-1/2 &  0 	&\sqb{12}^3 \anb{23}\\
	& 3/2 & 1/2 & 1/2 &-1/2	&\sqb{12}^2 \sqb{13} \anb{24}\\
	& 3/2 & 1/2 &  0  &  0 	&\sqb{12}^2 \sqb{13} \anb{23}\\
\text{dim-8}
	& 3/2 & 3/2 &-3/2 &-1/2	&\sqb{12}^4 \anb{13} \anb{23} \anb{34}\\
	& 3/2 & 3/2 & -1  &  0 	&\sqb{12}^4 \anb{13} \anb{23}\\
	& 3/2 &-3/2 &  1  & -1 	&\sqb{13}^3 \anb{23} \anb{24}^2\\
	& 3/2 &-3/2 & 1/2 &-1/2	&\sqb{13}^3 \anb{23}^2 \anb{24}\\
	& 3/2 &-3/2 &  0  &  0 	&\sqb{13}^3 \anb{23}^3\\
	& 3/2 &  1  &  1  &-1/2	&\sqb{12}^2 \sqb{13} \sqb{23} \anb{24}\\
	& 3/2 &  1  & -1  &-1/2	&\sqb{12}^3 \anb{23} \anb{34}\\
	& 3/2 & -1  & 1/2 &  0 	&\sqb{13}^3 \anb{23}^2\\
	& 3/2 & 1/2 &-1/2 &-1/2	&\sqb{12}^2 \sqb{13} \anb{23} \anb{34}\\
	& 3/2 &-1/2 &  0  &  0 	&\sqb{12} \sqb{13}^2 \anb{23}^2\\
\text{dim-9}
	& 3/2 & 3/2 & 3/2 &-3/2	&\sqb{12}^3 \sqb{23}^3 \anb{24}^3\\
	& 3/2 & 3/2 &-3/2 & 1/2	&\sqb{12}^4 \anb{13} \anb{23}^2 \sqb{24}\\
	& 3/2 & 3/2 &  1  & -1 	&\sqb{12}^3 \sqb{23}^2 \anb{24}^2\\
	& 3/2 &-3/2 &  1  &  0 	&\anb{12} \sqb{13}^4 \anb{23}^2\\
	& 3/2 &-3/2 & 1/2 & 1/2	&\sqb{13}^3 \anb{23}^3 \sqb{34}\\
	& 3/2 &  1  & -1  & 1/2	&\sqb{12}^3 \anb{23}^2 \sqb{24}\\
	& 3/2 & -1  & -1  & 1/2	&\sqb{12} \sqb{13} \sqb{14} \anb{23}^3\\
	& 3/2 & -1  &-1/2 &  0 	&\sqb{12} \sqb{13}^2 \anb{23}^3\\
	& 3/2 &-1/2 &-1/2 &-1/2	&\sqb{12} \sqb{13}^2 \anb{23}^2 \anb{34}\\
	&-3/2 &  1  &  1  &-1/2	&\anb{12} \anb{13} \anb{14} \sqb{23}^3\\
	&-3/2 &  1  & 1/2 &  0 	&\anb{12} \anb{13}^2 \sqb{23}^3\\
	&-3/2 & 1/2 & 1/2 & 1/2	&\anb{12} \anb{13}^2 \sqb{23}^2 \sqb{34}\\
\text{dim-10}
	& 3/2 &-3/2 &  1  &  1 	&\sqb{13}^3 \anb{23}^3 \sqb{34}^2\\
	&-3/2 &  1  &  1  & 1/2	&\anb{12} \anb{13}^2 \sqb{23}^3 \sqb{34}

%% file: 4pt-massless-two.tex
\text{dim-6}
	&  2  &  2  &  0  &  0 	&\sqb{12}^4\\
	&  2  & 3/2 & 3/2 &  0 	&\sqb{12}^2 \sqb{13}^2 \sqb{23}\\
	&  2  & 3/2 & 1/2 &  0 	&\sqb{12}^3 \sqb{13}\\
\text{dim-7}
	&  2  &  2  &  2  &  0 	&\sqb{12}^2 \sqb{13}^2 \sqb{23}^2\\
	&  2  &  2  & 3/2 & 3/2	&\sqb{12}^3 \sqb{13} \sqb{24} \sqb{34}^2, \sqb{12}^4 \sqb{34}^3, \sqb{12}^2 \sqb{13}^2 \sqb{24}^2 \sqb{34}, \sqb{12} \sqb{13}^3 \sqb{24}^3\\
	&  2  &  2  & 3/2 & 1/2	&\sqb{12}^3 \sqb{13} \sqb{23} \sqb{34}, \sqb{12}^2 \sqb{13}^2 \sqb{23} \sqb{24}\\
	&  2  &  2  &-3/2 &-3/2	&\sqb{12}^4 \anb{34}^3\\
	&  2  &  2  &  1  &  0 	&\sqb{12}^3 \sqb{13} \sqb{23}\\
	&  2  &  2  & 1/2 & 1/2	&\sqb{12}^3 \sqb{13} \sqb{24}, \sqb{12}^4 \sqb{34}\\
	&  2  &  2  &-1/2 &-1/2	&\sqb{12}^4 \anb{34}\\
	&  2  & 3/2 & 3/2 &  1 	&\sqb{12}^2 \sqb{13}^2 \sqb{24} \sqb{34}, \sqb{12} \sqb{13}^3 \sqb{24}^2, \sqb{12}^3 \sqb{13} \sqb{34}^2\\
	&  2  & 3/2 &  1  & 1/2	&\sqb{12}^3 \sqb{13} \sqb{34}, \sqb{12}^2 \sqb{13}^2 \sqb{24}\\
	&  2  & 3/2 &-1/2 &  0 	&\sqb{12}^4 \anb{23}\\
	&  2  &  1  &  1  &  0 	&\sqb{12}^2 \sqb{13}^2\\
	&  2  &  1  & 1/2 & 1/2	&\sqb{12}^2 \sqb{13} \sqb{14}\\
\text{dim-8}
	&  2  &  2  &  2  &  2 	&\sqb{13}^4 \sqb{24}^4, \sqb{12}^2 \sqb{13}^2 \sqb{24}^2 \sqb{34}^2, \sqb{12}^4 \sqb{34}^4, \sqb{12} \sqb{13}^3 \sqb{24}^3 \sqb{34}, \sqb{12}^3 \sqb{13} \sqb{24} \sqb{34}^3\\
	&  2  &  2  &  2  &  1 	&\sqb{12}^3 \sqb{13} \sqb{23} \sqb{34}^2, \sqb{12} \sqb{13}^3 \sqb{23} \sqb{24}^2, \sqb{12}^2 \sqb{13}^2 \sqb{23} \sqb{24} \sqb{34}\\
	&  2  &  2  & -2  & -2 	&\sqb{12}^4 \anb{34}^4\\
	&  2  &  2  & 3/2 &-1/2	&\sqb{12}^3 \sqb{13} \sqb{23}^2 \anb{24}\\
	&  2  &  2  &  1  &  1 	&\sqb{12}^3 \sqb{13} \sqb{24} \sqb{34}, \sqb{12}^2 \sqb{13}^2 \sqb{24}^2, \sqb{12}^4 \sqb{34}^2\\
	&  2  &  2  & -1  & -1 	&\sqb{12}^4 \anb{34}^2\\
	&  2  &  2  & 1/2 &-1/2	&\sqb{12}^4 \sqb{23} \anb{24}\\
	&  2  & -2  & 3/2 &-3/2	&\sqb{13}^4 \anb{23} \anb{24}^3\\
	&  2  & 3/2 &-3/2 & -1 	&\sqb{12}^4 \anb{23} \anb{34}^2\\
	&  2  & 3/2 &  1  &-1/2	&\sqb{12}^3 \sqb{13} \sqb{23} \anb{24}\\
	&  2  & 3/2 & -1  &-1/2	&\sqb{12}^4 \anb{23} \anb{34}\\
	&  2  &  1  &  1  &  1 	&\sqb{12}^2 \sqb{13} \sqb{14} \sqb{34}, \sqb{12} \sqb{13}^2 \sqb{14} \sqb{24}\\
	&  2  &  1  & 1/2 &-1/2	&\sqb{12}^3 \sqb{13} \anb{24}\\
	&  2  &  1  &  0  &  0 	&\sqb{12}^3 \sqb{13} \anb{23}\\
	&  2  & 1/2 & 1/2 &  0 	&\sqb{12}^2 \sqb{13}^2 \anb{23}\\
\text{dim-9}
	&  2  &  2  &-3/2 &-1/2	&\sqb{12}^5 \anb{13} \anb{23} \anb{34}\\
	&  2  &  2  & -1  &  0 	&\sqb{12}^5 \anb{13} \anb{23}\\
	&  2  & 3/2 & 3/2 & -1 	&\sqb{12}^3 \sqb{13} \sqb{23}^2 \anb{24}^2\\
	&  2  & 3/2 &-3/2 &  0 	&\sqb{12}^5 \anb{13} \anb{23}^2\\
	&  2  & 3/2 & -1  & 1/2	&\sqb{12}^4 \anb{23}^2 \sqb{24}\\
	&  2  &-3/2 &-3/2 &  1 	&\sqb{12} \sqb{13} \sqb{14}^2 \anb{23}^4\\
	&  2  &-3/2 &  1  &-1/2	&\sqb{13}^4 \anb{23}^2 \anb{24}\\
	&  2  &-3/2 & 1/2 &  0 	&\sqb{13}^4 \anb{23}^3\\
	&  2  &  1  & -1  &  0 	&\sqb{12}^4 \anb{23}^2\\
	&  2  &  1  &-1/2 &-1/2	&\sqb{12}^3 \sqb{13} \anb{23} \anb{34}\\
	&  2  & -1  & 1/2 & 1/2	&\sqb{13}^3 \sqb{14} \anb{23}^2\\
	&  2  & 1/2 &-1/2 &  0 	&\sqb{12}^3 \sqb{13} \anb{23}^2\\
	&  2  &  0  &  0  &  0 	&\sqb{12}^2 \sqb{13}^2 \anb{23}^2\\
	& -2  & 3/2 & 3/2 & -1 	&\anb{12} \anb{13} \anb{14}^2 \sqb{23}^4\\
\text{dim-10}
	&  2  &  2  &  2  & -1 	&\sqb{12}^3 \sqb{13} \sqb{23}^3 \anb{24}^2\\
	&  2  &  2  & -2  & -1 	&\sqb{12}^5 \anb{13} \anb{23} \anb{34}^2\\
	&  2  &  2  & 3/2 &-3/2	&\sqb{12}^4 \sqb{23}^3 \anb{24}^3\\
	&  2  &  2  &-3/2 & 1/2	&\sqb{12}^5 \anb{13} \anb{23}^2 \sqb{24}\\
	&  2  &  2  &  1  & -1 	&\sqb{12}^4 \sqb{23}^2 \anb{24}^2\\
	&  2  & -2  & 3/2 &-1/2	&\anb{12} \sqb{13}^5 \anb{23}^2 \anb{24}\\
	&  2  & -2  &  1  & -1 	&\sqb{13}^4 \anb{23}^2 \anb{24}^2\\
	&  2  & -2  & 1/2 &-1/2	&\sqb{13}^4 \anb{23}^3 \anb{24}\\
	&  2  & -2  &  0  &  0 	&\sqb{13}^4 \anb{23}^4\\
	&  2  & 3/2 &-3/2 &  1 	&\sqb{12}^4 \anb{23}^3 \sqb{24}^2\\
	&  2  &-3/2 &  1  & 1/2	&\sqb{13}^4 \anb{23}^3 \sqb{34}\\
	&  2  &-3/2 & -1  & 1/2	&\sqb{12} \sqb{13}^2 \sqb{14} \anb{23}^4\\
	&  2  &-3/2 &-1/2 &  0 	&\sqb{12} \sqb{13}^3 \anb{23}^4\\
	&  2  &  1  &  1  & -1 	&\sqb{12}^3 \sqb{13} \sqb{23} \anb{24}^2\\
	&  2  &  1  & -1  & -1 	&\sqb{12}^3 \sqb{13} \anb{23} \anb{34}^2\\
	&  2  & -1  & 1/2 &-1/2	&\sqb{12} \sqb{13}^3 \anb{23}^2 \anb{24}\\
	&  2  & -1  &  0  &  0 	&\sqb{12} \sqb{13}^3 \anb{23}^3\\
	&  2  &-1/2 &-1/2 &  0 	&\sqb{12}^2 \sqb{13}^2 \anb{23}^3\\
	& -2  & 3/2 & 3/2 &  0 	&\anb{12}^2 \anb{13}^2 \sqb{23}^5\\
	& -2  & 3/2 &  1  &-1/2	&\anb{12} \anb{13}^2 \anb{14} \sqb{23}^4\\
	& -2  & 3/2 & 1/2 &  0 	&\anb{12} \anb{13}^3 \sqb{23}^4\\
\text{dim-11}
	&  2  &  2  & -2  &  0 	&\sqb{12}^6 \anb{13}^2 \anb{23}^2\\
	&  2  & -2  & 3/2 & 3/2	&\sqb{13}^4 \anb{23}^4 \sqb{34}^3\\
	&  2  & -2  & 3/2 & 1/2	&\anb{12} \sqb{13}^5 \anb{23}^3 \sqb{34}\\
	&  2  & -2  &  1  &  0 	&\anb{12} \sqb{13}^5 \anb{23}^3\\
	&  2  & -2  & 1/2 & 1/2	&\sqb{13}^4 \anb{23}^4 \sqb{34}\\
	&  2  & -1  & -1  &  0 	&\sqb{12}^2 \sqb{13}^2 \anb{23}^4\\
	&  2  & -1  &-1/2 &-1/2	&\sqb{12} \sqb{13}^3 \anb{23}^3 \anb{34}\\
	& -2  & 3/2 & 3/2 &  1 	&\anb{12} \anb{13}^3 \sqb{23}^4 \sqb{34}^2\\
	& -2  & 3/2 &  1  & 1/2	&\anb{12} \anb{13}^3 \sqb{23}^4 \sqb{34}\\
	& -2  &  1  &  1  &  0 	&\anb{12}^2 \anb{13}^2 \sqb{23}^4\\
	& -2  &  1  & 1/2 & 1/2	&\anb{12} \anb{13}^3 \sqb{23}^3 \sqb{34}\\
\text{dim-12}
	&  2  &  2  &  2  & -2 	&\sqb{12}^4 \sqb{23}^4 \anb{24}^4\\
	&  2  &  2  & -2  &  1 	&\sqb{12}^5 \anb{13} \anb{23}^3 \sqb{24}^2\\
	&  2  & -2  &  1  &  1 	&\sqb{13}^4 \anb{23}^4 \sqb{34}^2\\
	& -2  &  1  &  1  &  1 	&\anb{12} \anb{13}^3 \sqb{23}^3 \sqb{34}^2

%% file: notes.bbl
\begin{thebibliography}{35}%
\makeatletter
\providecommand \@ifxundefined [1]{%
 \@ifx{#1\undefined}
}%
\providecommand \@ifnum [1]{%
 \ifnum #1\expandafter \@firstoftwo
 \else \expandafter \@secondoftwo
 \fi
}%
\providecommand \@ifx [1]{%
 \ifx #1\expandafter \@firstoftwo
 \else \expandafter \@secondoftwo
 \fi
}%
\providecommand \natexlab [1]{#1}%
\providecommand \enquote  [1]{``#1''}%
\providecommand \bibnamefont  [1]{#1}%
\providecommand \bibfnamefont [1]{#1}%
\providecommand \citenamefont [1]{#1}%
\providecommand \href@noop [0]{\@secondoftwo}%
\providecommand \href [0]{\begingroup \@sanitize@url \@href}%
\providecommand \@href[1]{\@@startlink{#1}\@@href}%
\providecommand \@@href[1]{\endgroup#1\@@endlink}%
\providecommand \@sanitize@url [0]{\catcode `\\12\catcode `\$12\catcode
  `\&12\catcode `\#12\catcode `\^12\catcode `\_12\catcode `\%12\relax}%
\providecommand \@@startlink[1]{}%
\providecommand \@@endlink[0]{}%
\providecommand \url  [0]{\begingroup\@sanitize@url \@url }%
\providecommand \@url [1]{\endgroup\@href {#1}{\urlprefix }}%
\providecommand \urlprefix  [0]{URL }%
\providecommand \Eprint [0]{\href }%
\providecommand \doibase [0]{http://dx.doi.org/}%
\providecommand \selectlanguage [0]{\@gobble}%
\providecommand \bibinfo [0]{\@secondoftwo}%
\providecommand \bibfield [0]{\@secondoftwo}%
\providecommand \translation [1]{[#1]}%
\providecommand \BibitemOpen [0]{}%
\providecommand \bibitemStop [0]{}%
\providecommand \bibitemNoStop [0]{.\EOS\space}%
\providecommand \EOS [0]{\spacefactor3000\relax}%
\providecommand \BibitemShut  [1]{\csname bibitem#1\endcsname}%
\let\auto@bib@innerbib\@empty
\bibitem [{\citenamefont{Cheung} and
  \citenamefont{Shen}(2015)}]{Cheung:2015aba}%
  \BibitemOpen
  \bibfield{author}{\bibinfo{author}{\bibfnamefont{C.}\,\bibnamefont{Cheung}}
  and \bibinfo{author}{\bibfnamefont{C.-H.} \bibnamefont{Shen}},
  }\bibfield{title}{\emph {\bibinfo{title}{{Nonrenormalization Theorems without
  Supersymmetry}}}, }\href {\doibase 10.1103/PhysRevLett.115.071601}
  {\bibfield{journal}{\bibinfo{journal}{Phys. Rev.
  Lett.}\,}\textbf{\bibinfo{volume}{115}}\,(\bibinfo{year}{2015})\,\bibinfo{pages}{071601}},
  \Eprint {http://arxiv.org/abs/1505.01844}{arXiv:1505.01844
  [hep-ph]}\BibitemShut {NoStop}%
\bibitem [{\citenamefont{Azatov} \emph {et\,al.}(2017)\citenamefont{Azatov},
  \citenamefont{Contino}, \citenamefont{Machado}, and
  \citenamefont{Riva}}]{Azatov:2016sqh}%
  \BibitemOpen
  \bibfield{author}{\bibinfo{author}{\bibfnamefont{A.}\,\bibnamefont{Azatov}},
  \bibinfo{author}{\bibfnamefont{R.}\,\bibnamefont{Contino}},
  \bibinfo{author}{\bibfnamefont{C.~S.} \bibnamefont{Machado}},  and
  \bibinfo{author}{\bibfnamefont{F.}\,\bibnamefont{Riva}},
  }\bibfield{title}{\emph {\bibinfo{title}{{Helicity selection rules and
  noninterference for BSM amplitudes}}}, }\href {\doibase
  10.1103/PhysRevD.95.065014} {\bibfield{journal}{\bibinfo{journal}{Phys.
  Rev.}\,}\textbf{\bibinfo{volume}{D95}}\,(\bibinfo{year}{2017})\,\bibinfo{pages}{065014}},
  \Eprint {http://arxiv.org/abs/1607.05236}{arXiv:1607.05236
  [hep-ph]}\BibitemShut {NoStop}%
\bibitem [{\citenamefont{Bern} \emph
  {et\,al.}(2020{\natexlab{a}})\citenamefont{Bern},
  \citenamefont{Parra-Martinez}, and \citenamefont{Sawyer}}]{Bern:2019wie}%
  \BibitemOpen
  \bibfield{author}{\bibinfo{author}{\bibfnamefont{Z.}\,\bibnamefont{Bern}},
  \bibinfo{author}{\bibfnamefont{J.}\,\bibnamefont{Parra-Martinez}},  and
  \bibinfo{author}{\bibfnamefont{E.}\,\bibnamefont{Sawyer}},
  }\bibfield{title}{\emph {\bibinfo{title}{{Nonrenormalization and Operator
  Mixing via On-Shell Methods}}}, }\href {\doibase
  10.1103/PhysRevLett.124.051601} {\bibfield{journal}{\bibinfo{journal}{Phys.
  Rev.
  Lett.}\,}\textbf{\bibinfo{volume}{124}}\,(\bibinfo{year}{2020}{\natexlab{a}})\,\bibinfo{pages}{051601}},
  \Eprint {http://arxiv.org/abs/1910.05831}{arXiv:1910.05831
  [hep-ph]}\BibitemShut {NoStop}%
\bibitem [{\citenamefont{Bern} \emph
  {et\,al.}(2020{\natexlab{b}})\citenamefont{Bern},
  \citenamefont{Parra-Martinez}, and \citenamefont{Sawyer}}]{Bern:2020ikv}%
  \BibitemOpen
  \bibfield{author}{\bibinfo{author}{\bibfnamefont{Z.}\,\bibnamefont{Bern}},
  \bibinfo{author}{\bibfnamefont{J.}\,\bibnamefont{Parra-Martinez}},  and
  \bibinfo{author}{\bibfnamefont{E.}\,\bibnamefont{Sawyer}},
  }\bibfield{title}{\emph {\bibinfo{title}{{Structure of two-loop SMEFT
  anomalous dimensions via on-shell methods}}}, }\href {\doibase
  10.1007/JHEP10(2020)211}
  {\bibfield{journal}{\bibinfo{journal}{JHEP}\,}\textbf{\bibinfo{volume}{10}}\,(\bibinfo{year}{2020}{\natexlab{b}})\,\bibinfo{pages}{211}},
  \Eprint {http://arxiv.org/abs/2005.12917}{arXiv:2005.12917
  [hep-ph]}\BibitemShut {NoStop}%
\bibitem [{\citenamefont{Jiang} \emph {et\,al.}(2021)\citenamefont{Jiang},
  \citenamefont{Ma}, and \citenamefont{Shu}}]{Jiang:2020mhe}%
  \BibitemOpen
  \bibfield{author}{\bibinfo{author}{\bibfnamefont{M.}\,\bibnamefont{Jiang}},
  \bibinfo{author}{\bibfnamefont{T.}\,\bibnamefont{Ma}},  and
  \bibinfo{author}{\bibfnamefont{J.}\,\bibnamefont{Shu}},
  }\bibfield{title}{\emph {\bibinfo{title}{{Renormalization Group Evolution
  from On-shell SMEFT}}}, }\href {\doibase 10.1007/JHEP01(2021)101}
  {\bibfield{journal}{\bibinfo{journal}{JHEP}\,}\textbf{\bibinfo{volume}{01}}\,(\bibinfo{year}{2021})\,\bibinfo{pages}{101}},
  \Eprint {http://arxiv.org/abs/2005.10261}{arXiv:2005.10261
  [hep-ph]}\BibitemShut {NoStop}%
\bibitem [{\citenamefont{Elias~Mir\'o} \emph
  {et\,al.}(2020)\citenamefont{Elias~Mir\'o}, \citenamefont{Ingoldby}, and
  \citenamefont{Riembau}}]{EliasMiro:2020tdv}%
  \BibitemOpen
  \bibfield{author}{\bibinfo{author}{\bibfnamefont{J.}\,\bibnamefont{Elias~Mir\'o}},
  \bibinfo{author}{\bibfnamefont{J.}\,\bibnamefont{Ingoldby}},  and
  \bibinfo{author}{\bibfnamefont{M.}\,\bibnamefont{Riembau}},
  }\bibfield{title}{\emph {\bibinfo{title}{{EFT anomalous dimensions from the
  S-matrix}}}, }\href {\doibase 10.1007/JHEP09(2020)163}
  {\bibfield{journal}{\bibinfo{journal}{JHEP}\,}\textbf{\bibinfo{volume}{09}}\,(\bibinfo{year}{2020})\,\bibinfo{pages}{163}},
  \Eprint {http://arxiv.org/abs/2005.06983}{arXiv:2005.06983
  [hep-ph]}\BibitemShut {NoStop}%
\bibitem [{\citenamefont{Arkani-Hamed} \emph
  {et\,al.}(2021)\citenamefont{Arkani-Hamed}, \citenamefont{Huang}, and
  \citenamefont{Huang}}]{Arkani-Hamed:2017jhn}%
  \BibitemOpen
  \bibfield{author}{\bibinfo{author}{\bibfnamefont{N.}\,\bibnamefont{Arkani-Hamed}},
  \bibinfo{author}{\bibfnamefont{T.-C.} \bibnamefont{Huang}},  and
  \bibinfo{author}{\bibfnamefont{Y.-t.} \bibnamefont{Huang}},
  }\bibfield{title}{\emph {\bibinfo{title}{{Scattering amplitudes for all
  masses and spins}}}, }\href {\doibase 10.1007/JHEP11(2021)070}
  {\bibfield{journal}{\bibinfo{journal}{JHEP}\,}\textbf{\bibinfo{volume}{11}}\,(\bibinfo{year}{2021})\,\bibinfo{pages}{070}},
  \Eprint {http://arxiv.org/abs/1709.04891}{arXiv:1709.04891
  [hep-th]}\BibitemShut {NoStop}%
\bibitem [{\citenamefont{Conde} and
  \citenamefont{Marzolla}(2016)}]{Conde:2016vxs}%
  \BibitemOpen
  \bibfield{author}{\bibinfo{author}{\bibfnamefont{E.}\,\bibnamefont{Conde}}
  and \bibinfo{author}{\bibfnamefont{A.}\,\bibnamefont{Marzolla}},
  }\bibfield{title}{\emph {\bibinfo{title}{{Lorentz Constraints on Massive
  Three-Point Amplitudes}}}, }\href {\doibase 10.1007/JHEP09(2016)041}
  {\bibfield{journal}{\bibinfo{journal}{JHEP}\,}\textbf{\bibinfo{volume}{09}}\,(\bibinfo{year}{2016})\,\bibinfo{pages}{041}},
  \Eprint {http://arxiv.org/abs/1601.08113}{arXiv:1601.08113
  [hep-th]}\BibitemShut {NoStop}%
\bibitem [{\citenamefont{Henning} \emph {et\,al.}(2016)\citenamefont{Henning},
  \citenamefont{Lu}, \citenamefont{Melia}, and
  \citenamefont{Murayama}}]{Henning:2015daa}%
  \BibitemOpen
  \bibfield{author}{\bibinfo{author}{\bibfnamefont{B.}\,\bibnamefont{Henning}},
  \bibinfo{author}{\bibfnamefont{X.}\,\bibnamefont{Lu}},
  \bibinfo{author}{\bibfnamefont{T.}\,\bibnamefont{Melia}},  and
  \bibinfo{author}{\bibfnamefont{H.}\,\bibnamefont{Murayama}},
  }\bibfield{title}{\emph {\bibinfo{title}{{Hilbert series and operator bases
  with derivatives in effective field theories}}}, }\href {\doibase
  10.1007/s00220-015-2518-2} {\bibfield{journal}{\bibinfo{journal}{Commun.
  Math.
  Phys.}\,}\textbf{\bibinfo{volume}{347}}\,(\bibinfo{year}{2016})\,\bibinfo{pages}{363}},
  \Eprint {http://arxiv.org/abs/1507.07240}{arXiv:1507.07240
  [hep-th]}\BibitemShut {NoStop}%
\bibitem [{\citenamefont{Henning} \emph {et\,al.}(2017)\citenamefont{Henning},
  \citenamefont{Lu}, \citenamefont{Melia}, and
  \citenamefont{Murayama}}]{Henning:2017fpj}%
  \BibitemOpen
  \bibfield{author}{\bibinfo{author}{\bibfnamefont{B.}\,\bibnamefont{Henning}},
  \bibinfo{author}{\bibfnamefont{X.}\,\bibnamefont{Lu}},
  \bibinfo{author}{\bibfnamefont{T.}\,\bibnamefont{Melia}},  and
  \bibinfo{author}{\bibfnamefont{H.}\,\bibnamefont{Murayama}},
  }\bibfield{title}{\emph {\bibinfo{title}{{Operator bases, $S$-matrices, and
  their partition functions}}}, }\href {\doibase 10.1007/JHEP10(2017)199}
  {\bibfield{journal}{\bibinfo{journal}{JHEP}\,}\textbf{\bibinfo{volume}{10}}\,(\bibinfo{year}{2017})\,\bibinfo{pages}{199}},
  \Eprint {http://arxiv.org/abs/1706.08520}{arXiv:1706.08520
  [hep-th]}\BibitemShut {NoStop}%
\bibitem [{\citenamefont{Shadmi} and
  \citenamefont{Weiss}(2019)}]{Shadmi:2018xan}%
  \BibitemOpen
  \bibfield{author}{\bibinfo{author}{\bibfnamefont{Y.}\,\bibnamefont{Shadmi}}
  and \bibinfo{author}{\bibfnamefont{Y.}\,\bibnamefont{Weiss}},
  }\bibfield{title}{\emph {\bibinfo{title}{{Effective Field Theory Amplitudes
  the On-Shell Way: Scalar and Vector Couplings to Gluons}}}, }\href {\doibase
  10.1007/JHEP02(2019)165}
  {\bibfield{journal}{\bibinfo{journal}{JHEP}\,}\textbf{\bibinfo{volume}{02}}\,(\bibinfo{year}{2019})\,\bibinfo{pages}{165}},
  \Eprint {http://arxiv.org/abs/1809.09644}{arXiv:1809.09644
  [hep-ph]}\BibitemShut {NoStop}%
\bibitem [{\citenamefont{Henning} and
  \citenamefont{Melia}(2019)}]{Henning:2019enq}%
  \BibitemOpen
  \bibfield{author}{\bibinfo{author}{\bibfnamefont{B.}\,\bibnamefont{Henning}}
  and \bibinfo{author}{\bibfnamefont{T.}\,\bibnamefont{Melia}},
  }\bibfield{title}{\emph {\bibinfo{title}{{Constructing effective field
  theories via their harmonics}}}, }\href {\doibase
  10.1103/PhysRevD.100.016015} {\bibfield{journal}{\bibinfo{journal}{Phys.
  Rev.}\,}\textbf{\bibinfo{volume}{D100}}\,(\bibinfo{year}{2019})\,\bibinfo{pages}{016015}},
  \Eprint {http://arxiv.org/abs/1902.06754}{arXiv:1902.06754
  [hep-ph]}\BibitemShut {NoStop}%
\bibitem [{\citenamefont{Henning} and
  \citenamefont{Melia}()}]{Henning:2019mcv}%
  \BibitemOpen
  \bibfield{author}{\bibinfo{author}{\bibfnamefont{B.}\,\bibnamefont{Henning}}
  and \bibinfo{author}{\bibfnamefont{T.}\,\bibnamefont{Melia}},
  }\bibfield{title}{\emph {\bibinfo{title}{{Conformal-helicity duality \& the
  Hilbert space of free CFTs}}}, }\href@noop {} {}\,\Eprint
  {http://arxiv.org/abs/1902.06747}{arXiv:1902.06747 [hep-th]}\BibitemShut
  {NoStop}%
\bibitem [{\citenamefont{Aoude} and
  \citenamefont{Machado}(2019)}]{Aoude:2019tzn}%
  \BibitemOpen
  \bibfield{author}{\bibinfo{author}{\bibfnamefont{R.}\,\bibnamefont{Aoude}}
  and \bibinfo{author}{\bibfnamefont{C.~S.} \bibnamefont{Machado}},
  }\bibfield{title}{\emph {\bibinfo{title}{{The Rise of SMEFT On-shell
  Amplitudes}}}, }\href {\doibase 10.1007/JHEP12(2019)058}
  {\bibfield{journal}{\bibinfo{journal}{JHEP}\,}\textbf{\bibinfo{volume}{12}}\,(\bibinfo{year}{2019})\,\bibinfo{pages}{058}},
  \Eprint {http://arxiv.org/abs/1905.11433}{arXiv:1905.11433
  [hep-ph]}\BibitemShut {NoStop}%
\bibitem [{\citenamefont{Durieux} \emph {et\,al.}(2020)\citenamefont{Durieux},
  \citenamefont{Kitahara}, \citenamefont{Shadmi}, and
  \citenamefont{Weiss}}]{Durieux:2019eor}%
  \BibitemOpen
  \bibfield{author}{\bibinfo{author}{\bibfnamefont{G.}\,\bibnamefont{Durieux}},
  \bibinfo{author}{\bibfnamefont{T.}\,\bibnamefont{Kitahara}},
  \bibinfo{author}{\bibfnamefont{Y.}\,\bibnamefont{Shadmi}},  and
  \bibinfo{author}{\bibfnamefont{Y.}\,\bibnamefont{Weiss}},
  }\bibfield{title}{\emph {\bibinfo{title}{{The electroweak effective field
  theory from on-shell amplitudes}}}, }\href {\doibase 10.1007/JHEP01(2020)119}
  {\bibfield{journal}{\bibinfo{journal}{JHEP}\,}\textbf{\bibinfo{volume}{01}}\,(\bibinfo{year}{2020})\,\bibinfo{pages}{119}},
  \Eprint {http://arxiv.org/abs/1909.10551}{arXiv:1909.10551
  [hep-ph]}\BibitemShut {NoStop}%
\bibitem [{\citenamefont{Durieux} and
  \citenamefont{Machado}(2020)}]{Durieux:2019siw}%
  \BibitemOpen
  \bibfield{author}{\bibinfo{author}{\bibfnamefont{G.}\,\bibnamefont{Durieux}}
  and \bibinfo{author}{\bibfnamefont{C.~S.} \bibnamefont{Machado}},
  }\bibfield{title}{\emph {\bibinfo{title}{{Enumerating higher-dimensional
  operators with on-shell amplitudes}}}, }\href {\doibase
  10.1103/PhysRevD.101.095021} {\bibfield{journal}{\bibinfo{journal}{Phys.
  Rev.}\,}\textbf{\bibinfo{volume}{D101}}\,(\bibinfo{year}{2020})\,\bibinfo{pages}{095021}},
  \Eprint {http://arxiv.org/abs/1912.08827}{arXiv:1912.08827
  [hep-ph]}\BibitemShut {NoStop}%
\bibitem [{\citenamefont{Ma} \emph {et\,al.}()\citenamefont{Ma},
  \citenamefont{Shu}, and \citenamefont{Xiao}}]{Ma:2019gtx}%
  \BibitemOpen
  \bibfield{author}{\bibinfo{author}{\bibfnamefont{T.}\,\bibnamefont{Ma}},
  \bibinfo{author}{\bibfnamefont{J.}\,\bibnamefont{Shu}},  and
  \bibinfo{author}{\bibfnamefont{M.-L.} \bibnamefont{Xiao}},
  }\bibfield{title}{\emph {\bibinfo{title}{{Standard Model Effective Field
  Theory from On-shell Amplitudes}}}, }\href@noop {} {}\,\Eprint
  {http://arxiv.org/abs/1902.06752}{arXiv:1902.06752 [hep-ph]}\BibitemShut
  {NoStop}%
\bibitem [{\citenamefont{Falkowski}()}]{Falkowski:2019zdo}%
  \BibitemOpen
  \bibfield{author}{\bibinfo{author}{\bibfnamefont{A.}\,\bibnamefont{Falkowski}},
  }\bibfield{title}{\emph {\bibinfo{title}{{Bases of massless EFTs via momentum
  twistors}}}, }\href@noop {} {}\,\Eprint
  {http://arxiv.org/abs/1912.07865}{arXiv:1912.07865 [hep-ph]}\BibitemShut
  {NoStop}%
\bibitem [{\citenamefont{Bern} \emph {et\,al.}(1994)\citenamefont{Bern},
  \citenamefont{Dixon}, \citenamefont{Dunbar}, and
  \citenamefont{Kosower}}]{Bern:1994zx}%
  \BibitemOpen
  \bibfield{author}{\bibinfo{author}{\bibfnamefont{Z.}\,\bibnamefont{Bern}},
  \bibinfo{author}{\bibfnamefont{L.~J.} \bibnamefont{Dixon}},
  \bibinfo{author}{\bibfnamefont{D.~C.} \bibnamefont{Dunbar}},  and
  \bibinfo{author}{\bibfnamefont{D.~A.} \bibnamefont{Kosower}},
  }\bibfield{title}{\emph {\bibinfo{title}{{One loop n point gauge theory
  amplitudes, unitarity and collinear limits}}}, }\href {\doibase
  10.1016/0550-3213(94)90179-1} {\bibfield{journal}{\bibinfo{journal}{Nucl.
  Phys.
  B}\,}\textbf{\bibinfo{volume}{425}}\,(\bibinfo{year}{1994})\,\bibinfo{pages}{217}},
  \Eprint
  {http://arxiv.org/abs/hep-ph/9403226}{arXiv:hep-ph/9403226}\BibitemShut
  {NoStop}%
\bibitem [{\citenamefont{Cohen} \emph {et\,al.}(2011)\citenamefont{Cohen},
  \citenamefont{Elvang}, and \citenamefont{Kiermaier}}]{Cohen:2010mi}%
  \BibitemOpen
  \bibfield{author}{\bibinfo{author}{\bibfnamefont{T.}\,\bibnamefont{Cohen}},
  \bibinfo{author}{\bibfnamefont{H.}\,\bibnamefont{Elvang}},  and
  \bibinfo{author}{\bibfnamefont{M.}\,\bibnamefont{Kiermaier}},
  }\bibfield{title}{\emph {\bibinfo{title}{{On-shell constructibility of tree
  amplitudes in general field theories}}}, }\href {\doibase
  10.1007/JHEP04(2011)053}
  {\bibfield{journal}{\bibinfo{journal}{JHEP}\,}\textbf{\bibinfo{volume}{04}}\,(\bibinfo{year}{2011})\,\bibinfo{pages}{053}},
  \Eprint {http://arxiv.org/abs/1010.0257}{arXiv:1010.0257
  [hep-th]}\BibitemShut {NoStop}%
\bibitem [{\citenamefont{Franken} and
  \citenamefont{Schwinn}(2020)}]{Franken:2019wqr}%
  \BibitemOpen
  \bibfield{author}{\bibinfo{author}{\bibfnamefont{R.}\,\bibnamefont{Franken}}
  and \bibinfo{author}{\bibfnamefont{C.}\,\bibnamefont{Schwinn}},
  }\bibfield{title}{\emph {\bibinfo{title}{{On-shell constructibility of Born
  amplitudes in spontaneously broken gauge theories}}}, }\href {\doibase
  10.1007/JHEP02(2020)073}
  {\bibfield{journal}{\bibinfo{journal}{JHEP}\,}\textbf{\bibinfo{volume}{02}}\,(\bibinfo{year}{2020})\,\bibinfo{pages}{073}},
  \Eprint {http://arxiv.org/abs/1910.13407}{arXiv:1910.13407
  [hep-th]}\BibitemShut {NoStop}%
\bibitem [{\citenamefont{Falkowski} and
  \citenamefont{Machado}(2021)}]{Falkowski:2020aso}%
  \BibitemOpen
  \bibfield{author}{\bibinfo{author}{\bibfnamefont{A.}\,\bibnamefont{Falkowski}}
  and \bibinfo{author}{\bibfnamefont{C.~S.} \bibnamefont{Machado}},
  }\bibfield{title}{\emph {\bibinfo{title}{{Soft Matters, or the Recursions
  with Massive Spinors}}}, }\href {\doibase 10.1007/JHEP05(2021)238}
  {\bibfield{journal}{\bibinfo{journal}{JHEP}\,}\textbf{\bibinfo{volume}{05}}\,(\bibinfo{year}{2021})\,\bibinfo{pages}{238}},
  \Eprint {http://arxiv.org/abs/2005.08981}{arXiv:2005.08981
  [hep-th]}\BibitemShut {NoStop}%
\bibitem [{\citenamefont{Bachu} and
  \citenamefont{Yelleshpur}(2020)}]{Bachu:2019ehv}%
  \BibitemOpen
  \bibfield{author}{\bibinfo{author}{\bibfnamefont{B.}\,\bibnamefont{Bachu}}
  and \bibinfo{author}{\bibfnamefont{A.}\,\bibnamefont{Yelleshpur}},
  }\bibfield{title}{\emph {\bibinfo{title}{{On-Shell Electroweak Sector and the
  Higgs Mechanism}}}, }\href {\doibase 10.1007/JHEP08(2020)039}
  {\bibfield{journal}{\bibinfo{journal}{JHEP}\,}\textbf{\bibinfo{volume}{08}}\,(\bibinfo{year}{2020})\,\bibinfo{pages}{039}},
  \Eprint {http://arxiv.org/abs/1912.04334}{arXiv:1912.04334
  [hep-th]}\BibitemShut {NoStop}%
\bibitem [{\citenamefont{Christensen} and
  \citenamefont{Field}(2018)}]{Christensen:2018zcq}%
  \BibitemOpen
  \bibfield{author}{\bibinfo{author}{\bibfnamefont{N.}\,\bibnamefont{Christensen}}
  and \bibinfo{author}{\bibfnamefont{B.}\,\bibnamefont{Field}},
  }\bibfield{title}{\emph {\bibinfo{title}{{Constructive standard model}}},
  }\href {\doibase 10.1103/PhysRevD.98.016014}
  {\bibfield{journal}{\bibinfo{journal}{Phys.
  Rev.}\,}\textbf{\bibinfo{volume}{D98}}\,(\bibinfo{year}{2018})\,\bibinfo{pages}{016014}},
  \Eprint {http://arxiv.org/abs/1802.00448}{arXiv:1802.00448
  [hep-ph]}\BibitemShut {NoStop}%
\bibitem [{\citenamefont{Alonso} \emph
  {et\,al.}(2016{\natexlab{a}})\citenamefont{Alonso}, \citenamefont{Jenkins},
  and \citenamefont{Manohar}}]{Alonso:2015fsp}%
  \BibitemOpen
  \bibfield{author}{\bibinfo{author}{\bibfnamefont{R.}\,\bibnamefont{Alonso}},
  \bibinfo{author}{\bibfnamefont{E.~E.} \bibnamefont{Jenkins}},  and
  \bibinfo{author}{\bibfnamefont{A.~V.} \bibnamefont{Manohar}},
  }\bibfield{title}{\emph {\bibinfo{title}{{A Geometric Formulation of Higgs
  Effective Field Theory: Measuring the Curvature of Scalar Field Space}}},
  }\href {\doibase 10.1016/j.physletb.2016.01.041}
  {\bibfield{journal}{\bibinfo{journal}{Phys.
  Lett.}\,}\textbf{\bibinfo{volume}{B754}}\,(\bibinfo{year}{2016}{\natexlab{a}})\,\bibinfo{pages}{335}},
  \Eprint {http://arxiv.org/abs/1511.00724}{arXiv:1511.00724
  [hep-ph]}\BibitemShut {NoStop}%
\bibitem [{\citenamefont{Alonso} \emph
  {et\,al.}(2016{\natexlab{b}})\citenamefont{Alonso}, \citenamefont{Jenkins},
  and \citenamefont{Manohar}}]{Alonso:2016oah}%
  \BibitemOpen
  \bibfield{author}{\bibinfo{author}{\bibfnamefont{R.}\,\bibnamefont{Alonso}},
  \bibinfo{author}{\bibfnamefont{E.~E.} \bibnamefont{Jenkins}},  and
  \bibinfo{author}{\bibfnamefont{A.~V.} \bibnamefont{Manohar}},
  }\bibfield{title}{\emph {\bibinfo{title}{{Geometry of the Scalar Sector}}},
  }\href {\doibase 10.1007/JHEP08(2016)101}
  {\bibfield{journal}{\bibinfo{journal}{JHEP}\,}\textbf{\bibinfo{volume}{08}}\,(\bibinfo{year}{2016}{\natexlab{b}})\,\bibinfo{pages}{101}},
  \Eprint {http://arxiv.org/abs/1605.03602}{arXiv:1605.03602
  [hep-ph]}\BibitemShut {NoStop}%
\bibitem [{\citenamefont{Helset} \emph {et\,al.}(2020)\citenamefont{Helset},
  \citenamefont{Martin}, and \citenamefont{Trott}}]{Helset:2020yio}%
  \BibitemOpen
  \bibfield{author}{\bibinfo{author}{\bibfnamefont{A.}\,\bibnamefont{Helset}},
  \bibinfo{author}{\bibfnamefont{A.}\,\bibnamefont{Martin}},  and
  \bibinfo{author}{\bibfnamefont{M.}\,\bibnamefont{Trott}},
  }\bibfield{title}{\emph {\bibinfo{title}{{The Geometric Standard Model
  Effective Field Theory}}}, }\href {\doibase 10.1007/JHEP03(2020)163}
  {\bibfield{journal}{\bibinfo{journal}{JHEP}\,}\textbf{\bibinfo{volume}{03}}\,(\bibinfo{year}{2020})\,\bibinfo{pages}{163}},
  \Eprint {http://arxiv.org/abs/2001.01453}{arXiv:2001.01453
  [hep-ph]}\BibitemShut {NoStop}%
\bibitem [{\citenamefont{Cohen} \emph {et\,al.}(2021)\citenamefont{Cohen},
  \citenamefont{Craig}, \citenamefont{Lu}, and
  \citenamefont{Sutherland}}]{Cohen:2020xca}%
  \BibitemOpen
  \bibfield{author}{\bibinfo{author}{\bibfnamefont{T.}\,\bibnamefont{Cohen}},
  \bibinfo{author}{\bibfnamefont{N.}\,\bibnamefont{Craig}},
  \bibinfo{author}{\bibfnamefont{X.}\,\bibnamefont{Lu}},  and
  \bibinfo{author}{\bibfnamefont{D.}\,\bibnamefont{Sutherland}},
  }\bibfield{title}{\emph {\bibinfo{title}{{Is SMEFT Enough?}}}, }\href
  {\doibase 10.1007/JHEP03(2021)237}
  {\bibfield{journal}{\bibinfo{journal}{JHEP}\,}\textbf{\bibinfo{volume}{03}}\,(\bibinfo{year}{2021})\,\bibinfo{pages}{237}},
  \Eprint {http://arxiv.org/abs/2008.08597}{arXiv:2008.08597
  [hep-ph]}\BibitemShut {NoStop}%
\bibitem [{\citenamefont{Costa} \emph {et\,al.}(2011)\citenamefont{Costa},
  \citenamefont{Penedones}, \citenamefont{Poland}, and
  \citenamefont{Rychkov}}]{Costa:2011mg}%
  \BibitemOpen
  \bibfield{author}{\bibinfo{author}{\bibfnamefont{M.~S.} \bibnamefont{Costa}},
  \bibinfo{author}{\bibfnamefont{J.}\,\bibnamefont{Penedones}},
  \bibinfo{author}{\bibfnamefont{D.}\,\bibnamefont{Poland}},  and
  \bibinfo{author}{\bibfnamefont{S.}\,\bibnamefont{Rychkov}},
  }\bibfield{title}{\emph {\bibinfo{title}{{Spinning Conformal Correlators}}},
  }\href {\doibase 10.1007/JHEP11(2011)071}
  {\bibfield{journal}{\bibinfo{journal}{JHEP}\,}\textbf{\bibinfo{volume}{11}}\,(\bibinfo{year}{2011})\,\bibinfo{pages}{071}},
  \Eprint {http://arxiv.org/abs/1107.3554}{arXiv:1107.3554
  [hep-th]}\BibitemShut {NoStop}%
\bibitem [{\citenamefont{Mangano} and
  \citenamefont{Parke}(1991)}]{Mangano:1990by}%
  \BibitemOpen
  \bibfield{author}{\bibinfo{author}{\bibfnamefont{M.~L.}
  \bibnamefont{Mangano}} and \bibinfo{author}{\bibfnamefont{S.~J.}
  \bibnamefont{Parke}}, }\bibfield{title}{\emph {\bibinfo{title}{{Multiparton
  amplitudes in gauge theories}}}, }\href {\doibase
  10.1016/0370-1573(91)90091-Y} {\bibfield{journal}{\bibinfo{journal}{Phys.
  Rept.}\,}\textbf{\bibinfo{volume}{200}}\,(\bibinfo{year}{1991})\,\bibinfo{pages}{301}},
  \Eprint
  {http://arxiv.org/abs/hep-th/0509223}{arXiv:hep-th/0509223}\BibitemShut
  {NoStop}%
\bibitem [{\citenamefont{Schomerus} \emph
  {et\,al.}(2017)\citenamefont{Schomerus}, \citenamefont{Sobko}, and
  \citenamefont{Isachenkov}}]{Schomerus:2016epl}%
  \BibitemOpen
  \bibfield{author}{\bibinfo{author}{\bibfnamefont{V.}\,\bibnamefont{Schomerus}},
  \bibinfo{author}{\bibfnamefont{E.}\,\bibnamefont{Sobko}},  and
  \bibinfo{author}{\bibfnamefont{M.}\,\bibnamefont{Isachenkov}},
  }\bibfield{title}{\emph {\bibinfo{title}{{Harmony of Spinning Conformal
  Blocks}}}, }\href {\doibase 10.1007/JHEP03(2017)085}
  {\bibfield{journal}{\bibinfo{journal}{JHEP}\,}\textbf{\bibinfo{volume}{03}}\,(\bibinfo{year}{2017})\,\bibinfo{pages}{085}},
  \Eprint {http://arxiv.org/abs/1612.02479}{arXiv:1612.02479
  [hep-th]}\BibitemShut {NoStop}%
\bibitem [{\citenamefont{Kravchuk} and
  \citenamefont{Simmons-Duffin}(2018)}]{Kravchuk:2016qvl}%
  \BibitemOpen
  \bibfield{author}{\bibinfo{author}{\bibfnamefont{P.}\,\bibnamefont{Kravchuk}}
  and \bibinfo{author}{\bibfnamefont{D.}\,\bibnamefont{Simmons-Duffin}},
  }\bibfield{title}{\emph {\bibinfo{title}{{Counting Conformal Correlators}}},
  }\href {\doibase 10.1007/JHEP02(2018)096}
  {\bibfield{journal}{\bibinfo{journal}{JHEP}\,}\textbf{\bibinfo{volume}{02}}\,(\bibinfo{year}{2018})\,\bibinfo{pages}{096}},
  \Eprint {http://arxiv.org/abs/1612.08987}{arXiv:1612.08987
  [hep-th]}\BibitemShut {NoStop}%
\bibitem [{\citenamefont{Bonifacio} and
  \citenamefont{Hinterbichler}(2018)}]{Bonifacio:2018vzv}%
  \BibitemOpen
  \bibfield{author}{\bibinfo{author}{\bibfnamefont{J.}\,\bibnamefont{Bonifacio}}
  and \bibinfo{author}{\bibfnamefont{K.}\,\bibnamefont{Hinterbichler}},
  }\bibfield{title}{\emph {\bibinfo{title}{{Bounds on Amplitudes in Effective
  Theories with Massive Spinning Particles}}}, }\href {\doibase
  10.1103/PhysRevD.98.045003} {\bibfield{journal}{\bibinfo{journal}{Phys.
  Rev.}\,}\textbf{\bibinfo{volume}{D98}}\,(\bibinfo{year}{2018})\,\bibinfo{pages}{045003}},
  \Eprint {http://arxiv.org/abs/1804.08686}{arXiv:1804.08686
  [hep-th]}\BibitemShut {NoStop}%
\bibitem [{\citenamefont{Boels} and \citenamefont{Luo}(2018)}]{Boels:2017gyc}%
  \BibitemOpen
  \bibfield{author}{\bibinfo{author}{\bibfnamefont{R.~H.} \bibnamefont{Boels}}
  and \bibinfo{author}{\bibfnamefont{H.}\,\bibnamefont{Luo}},
  }\bibfield{title}{\emph {\bibinfo{title}{{A minimal approach to the
  scattering of physical massless bosons}}}, }\href {\doibase
  10.1007/JHEP05(2018)063}
  {\bibfield{journal}{\bibinfo{journal}{JHEP}\,}\textbf{\bibinfo{volume}{05}}\,(\bibinfo{year}{2018})\,\bibinfo{pages}{063}},
  \Eprint {http://arxiv.org/abs/1710.10208}{arXiv:1710.10208
  [hep-th]}\BibitemShut {NoStop}%
\bibitem [{\citenamefont{Glover} and
  \citenamefont{Tejeda-Yeomans}(2003)}]{Glover:2003cm}%
  \BibitemOpen
  \bibfield{author}{\bibinfo{author}{\bibfnamefont{E.~W.~N.}
  \bibnamefont{Glover}} and \bibinfo{author}{\bibfnamefont{M.~E.}
  \bibnamefont{Tejeda-Yeomans}}, }\bibfield{title}{\emph {\bibinfo{title}{{Two
  loop QCD helicity amplitudes for massless quark massless gauge boson
  scattering}}}, }\href {\doibase 10.1088/1126-6708/2003/06/033}
  {\bibfield{journal}{\bibinfo{journal}{JHEP}\,}\textbf{\bibinfo{volume}{06}}\,(\bibinfo{year}{2003})\,\bibinfo{pages}{033}},
  \Eprint
  {http://arxiv.org/abs/hep-ph/0304169}{arXiv:hep-ph/0304169}\BibitemShut
  {NoStop}%
\end{thebibliography}%
